\documentclass[preprint,12pt,authoryear]{elsarticle}
\usepackage{amssymb,amsfonts,amsthm}
\usepackage{amsbsy}
\usepackage{mathrsfs}
\usepackage{amsthm}
\usepackage{color,graphics,subfigure}
\usepackage{pifont}
\usepackage{natbib}
\usepackage{lineno}
\usepackage{bm}
\usepackage{setspace}
\usepackage{cite}
\usepackage{latexsym,multicol}
\usepackage{tcolorbox}
\usepackage[fleqn]{amsmath}
\numberwithin{equation}{section}
\usepackage{amssymb,amsfonts,amsthm}
\usepackage{geometry}
\usepackage{graphics,subfigure}
\usepackage{hyperref}
\usepackage{color}
\usepackage{lineno}
\usepackage{bm}
\newcommand{\gammacr}{\gamma_{\rm cr}}

\def\e{{\bm e}}

\def \div{\mbox{div\hskip 1pt}}

\def \tr{\mbox{tr\hskip 1pt}}

\def \tr{\operatorname{tr}}

\journal{Journal of the Mechanics and Physics of Solids}

\begin{document}
\numberwithin{equation}{section}
\begin{frontmatter}
\title{Tensional wrinkling of thin elastic sheets with two circular holes}

\author[add1,add2,add3]{Yang Liu}
\ead{tracy\_liu@tju.edu.cn} 
\author[add4,add5]{Sepideh Razavi}
\ead{srazavi@ou.edu}
\author[add4]{Pietro Cicuta}
\ead{pc245@cam.ac.uk}
\author[add2]{Dominic Vella}
\ead{dominic.vella@maths.ox.ac.uk}
\author[add2]{Alain Goriely\corref{cor1}}
\ead{alain.goriely@maths.ox.ac.uk} 
\address[add1]{Department of Mechanics, School of Mechanical Engineering, Tianjin University, Tianjin 300350, China}
\address[add2]{Mathematical Institute, University of Oxford, Oxford, OX2 6GG, UK}
\address[add3]{National Key Laboratory of Vehicle Power System, Tianjin 300350, China}
\address[add4]{Department of Physics, Cavendish Laboratory, University of Cambridge, Cambridge CB3 0HE, UK}
\address[add5]{School of Sustainable Chemical, Biological and Materials Engineering, University of Oklahoma, Oklahoma 73019, USA}

\cortext[cor1]{Corresponding author.}

\begin{abstract}
A paradigm for the study of wrinkling in elastic sheet is the  Lam\'{e} configuration, in which azimuthal wrinkles form in an annular sheet subjected to tensile loads at both edges. Since wrinkles are spatially extended, this instability provides a mechanism for stress transmission over long distances. A natural extension of this problem is  wrinkling in sheets with multiple holes or broken symmetry. Here, we investigate tension-induced wrinkling in thin elastic sheets containing two identical circular holes by combining analytical modeling and experiments. The pre-buckled state is solved analytically using bipolar coordinates within the framework of linear elasticity, enabling the identification of the wrinkling threshold as a function of the distance between the two holes. Near-threshold wrinkling and interactions between wrinkles are analyzed, and we validate our theoretical predictions against experimental observations obtained through video imaging of spin-coated polystyrene sheets floating on liquid surfaces with controlled surface tension. Our results demonstrate that geometric symmetry breaking, such as the presence of a second hole, strongly influences wrinkle nucleation, orientation, and spatial extent. Beyond mechanics, these findings further support the idea that mechanical cues can be amplified by instabilities, providing a mechanical analogue for long-range cellular mechanosensing. 

\end{abstract}

\begin{keyword}
		Ultrathin sheets \sep bipolar coordinates \sep elasticity \sep asymptotic analysis \sep capillary wrinkling
\end{keyword}

\end{frontmatter}

\section{Introduction}

Thin elastic sheets are much easier to bend than to stretch. Consequently, when subjected to compression \citep{bowden1998spontaneous,pocivavsek2008stress,liu2025post,liu2026brief}, poking \citep{vella2015indentation,dai2021poking,wang2023effects}, or being wrapped around a curved substrate \citep{hure2012stamping,hohlfeld2015sheet,box2023delamination}, they tend to deform out of plane by forming wrinkles, thereby relieving compressive stresses. A particularly important and subtle class of instabilities is tension-induced wrinkling, in which a sheet subjected to stretching forces develops localized compressive stresses due to force balance and compatibility constraints \citep{cerda2003geometry,wang2022mechanics,chai2024stretch}. A paradigm for such class of problems is the classical Lam\'{e} problem, consisting of a thin elastic annulus subjected to different radial tensions applied at its inner and outer boundaries. The primary deformation can be solved exactly using linear elasticity theory \citep{timoshenko}. Whereas  the radial principal stress is always tensile, the circumferential stress can become compressive, inducing wrinkling in a direction perpendicular to the compressive tractions \citep{coman2006localized,coman2007wrinkling}. A ground-breaking experiment by \citet{huang2007capillary} demonstrated the formation of ordered wrinkles when a liquid droplet is deposited on a thin polystyrene (PS) film floating on water. This observation led  to extensive research.

On the theoretical side, \citet{vella2010capillary} developed a physically based model and revealed that the presence of a liquid droplet generates azimuthal compression in the thin film, explaining the wrinkling reported in \citet{huang2007capillary}. It  also showed that the extent of the wrinkled region was underestimated. To resolve this discrepancy, tension-field theory was applied by \citet{davidovitch2011prototypical}, capturing the far-from-threshold wrinkling behavior. One of the most striking outcomes of such far-from-threshold analyses is the prediction that the spatial extent of wrinkles can be much larger than anticipated from near-threshold considerations. A refined analysis incorporating capillary-induced wrinkling was subsequently carried out by \citet{schroll2013capillary}, successfully explaining the wrinkle length observed experimentally by \citet{huang2007capillary}. Later, similar wrinkling phenomena were studied in annular geometries \citep{pineirua2013capillary,paulsen2017geometry}, in dynamic settings \citep{box2019dynamics}, as well as in the curved liquid sheets that encapsulate bubbles \citep{oratis2020new,davidovitch2024}.

The study of wrinkling in thin films is not only motivated by   the metrology of ultrathin films but also establishes connections with cellular mechanosensing and traction force microscopy for cells. In general, wrinkles effectively amplify weak compressive stresses, making them visible at macroscopic scales. From this perspective, wrinkling can be viewed as a mechanical analogue of signal propagation, in which information about force magnitude, direction, and geometry is encoded in the wrinkle morphology. In particular, along such wrinkles, cells can sense mechanical signals generated by other cells at a distance, which can regulate functions such as differentiation, proliferation, and apoptosis \citep{ladoux2017mechanobiology,vining2017mechanical,van2018mechanoreciprocity}. Cultured cells typically contract to produce stresses in the compliant substrate beneath them, which can be large enough to trigger elastic instabilities such as wrinkling. For instance, cells cultured on thin elastic membranes have been observed to generate wrinkle patterns aligned with actin stress fibers, with wrinkle orientation reflecting the anisotropy of cellular forces \citep{style2014traction}. Conversely,  wrinkle characteristics can be used to estimate cellular traction \citep{li2022wrinkle,Ardaseva2026}. The study of \citet{davidovitch2011prototypical} demonstrated that stress signals can be transmitted over longer distances in wrinkled regions --- this offers a qualitative analogy to cellular mechanosensing, where long-range mechanical cues can be sensed by surrounding cells to regulate collective cellular behaviors.   However, research so far has focused on an idealized single cell, neglecting possible mechanical interactions between multiple cells. As a first step,  it is natural to consider the case of two neighbouring cells, each generating a wrinkle field, but accounting for interactions. This motivates the present study, which considers a mechanical analogue of this problem.
 
Beyond this specific problem, we point out that in many realistic systems, symmetry is typically broken by geometric defects, inclusions, or boundaries, all of which lead to highly anisotropic stress fields. Perforated sheets, in particular, provide a fertile platform for exploring how geometry controls wrinkling patterns. \citet{paulsen2017geometry} experimentally studied the wrinkle-to-fold transition in a floating annulus subjected to inner and outer tensions, establishing a geometric rule for mode transitions and highlighting the role of geometry in regulating thin-film patterns. \citet{andrade2019pre} investigated a modified Lam\'{e} problem in which the circular hole is replaced by an elliptical one. It is worth mentioning that when geometric symmetry is further broken, for example, by adding an additional hole, abundant (and complex) wrinkle patterns can emerge, as shown in preliminary experiments with a lattice of droplets deposited  on a floating sheet \citep{huang2010wrinkling}. However, no theoretical model of this system has been presented so far. Since holes act as stress concentrators and introduce new length scales, the addition of another hole in the Lam\'{e} problem provides a rich system in which wrinkles may nucleate and interact in a manner distinct from the single-hole, axisymmetric case.

Here, we study tension-induced wrinkling in thin elastic sheets containing two identical circular holes --- what we term the bipolar Lam\'{e} problem. To do so, we use analytical techniques within the framework of linear elasticity to study the threshold and near-threshold behaviors as well as experiments to test these predictions and to start to indicate the behavior far from threshold. The remainder of the paper is organized as follows. In Section~\ref{Problem formulation}, we establish the mathematical formulation of the problem using bipolar coordinates. The exact solution for the pre-buckled state is derived in Section~\ref{Exact solution} using the technique of superposition. Section~\ref{onset} focuses on the onset of wrinkling, including obtaining asymptotic formulae for the critical tension ratio at which wrinkling is first observed. The near-threshold predictions for the spatial distribution of wrinkles are presented in Section~\ref{extent-of-wrinkles}, followed by the experimental investigation in Section~\ref{experiments}. Finally, discussions and conclusions are provided in Section~\ref{conclusion}.

\section{Problem formulation}\label{Problem formulation}
\begin{figure}[ht!]
	\centering\includegraphics[width=0.7\linewidth]{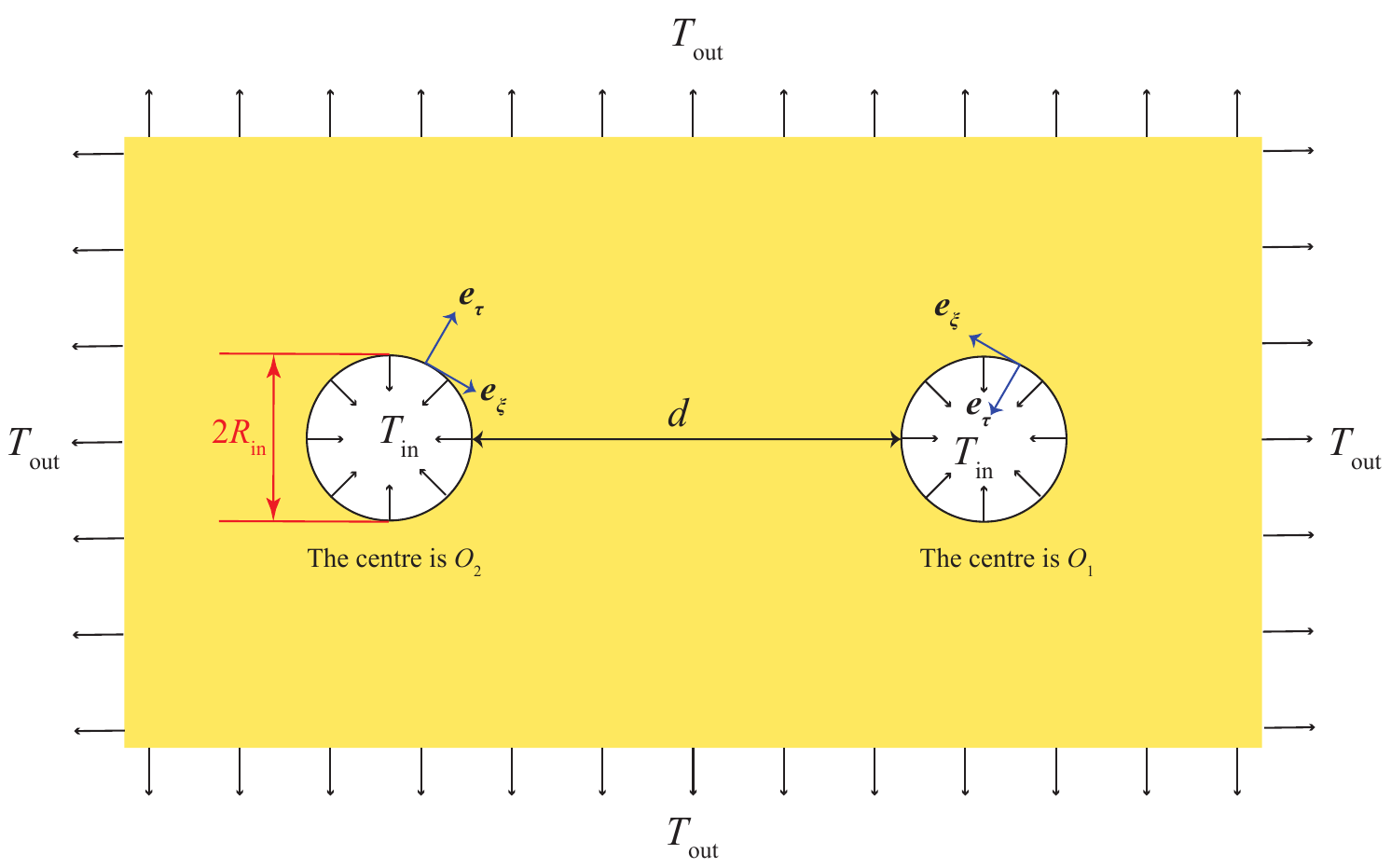}\caption{An infinite, elastic sheet (yellow) containing two equal circular holes of radius $R_\mathrm{in}$ and separated by a distance $d$ is subjected to a tensional stress $T_\mathrm{in}$ at the internal surfaces and an all-around tension $T_\mathrm{out}$ at infinity. }\label{fig:geometry}
\end{figure}

We consider a thin, linearly-elastic membrane with two circular holes separated by a distance $d$. The holes have the same radius $R_\mathrm{in}$ and their centers are denoted $O_1$ and $O_2$, respectively. The  membrane is subject to a uniform normal tension $T_\mathrm{in}$ along the boundary of the holes as well as an  isotropic tension $T_\mathrm{out}$ at infinity (Figure \ref{fig:geometry}).

\begin{figure}[htp]
    \centering
    \includegraphics[width=0.5\linewidth]{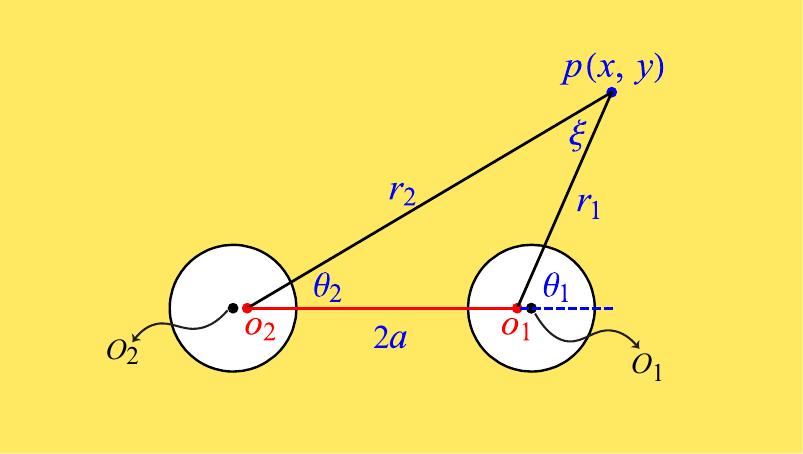}
    \caption{Definition of the bipolar coordinates $(\tau,\xi)$. The two foci are located at $o_1=(a,0)$ and $o_2=(-a,0)$ (red points). The angle $\xi=\angle o_1po_2$, while $\tau=\ln(r_2/r_1)$ is a measure of the relative distance from the point $p(x,y)$ of interest. The two black circles represent the locations of the holes in our problem with the corresponding centers highlighted by black points, and denoted $O_1$ and $O_2$, respectively.}
    \label{fig:bipolar}
\end{figure}

To  describe the deformation, we use the  bipolar coordinate system shown in Figure~\ref{fig:bipolar}; here bipolar coordinates $(\tau,\xi)$ are assigned to  a material point $p$ whose Cartesian coordinates $(x,y)$ satisfy:
\begin{equation}
x=\frac{a\sinh\tau}{\cosh\tau-\cos\xi},\quad y=\frac{a\sin\xi}{\cosh\tau-\cos\xi},\quad \tau\in\mathbb R,\quad\xi\in[0,\,2\pi].
\label{eq:coordinates}
\end{equation}
The two foci associated with the bipolar coordinates are denoted by $o_1$ and $o_2$, respectively, and the distance between them is $2a$. Geometrically, $\xi=\angle o_1po_2$ while $\tau$ relates the relative distance to each of the foci, since 
\begin{equation}
    \tau=\ln\frac{r_2}{r_1},
\end{equation} 
where  $r_i$ is the distance between $p$ and $o_i$. Note that at infinity, we have $\tau\rightarrow0$ and $\xi\rightarrow0$. 
 In addition, we define two angular variables $\theta_1$ and $\theta_2$, which will be employed in deriving asymptotic expressions.

We note that for constant $\xi$,
\begin{equation}
x^2+(y-a\cot\xi)^2=\frac{a^2}{\sin^2\xi},
\label{eq:circle-y}
\end{equation}
which is a circle with center $a\cot\xi$ and radius $a/|\sin\xi|$. Likewise, the contours of constant $\tau$ are characterized by
\begin{equation}
(x-a\coth\tau)^2+y^2=\frac{a^2}{\sinh^2\tau},
\label{eq:circle-x}
\end{equation} which are again circles.

In bipolar coordinates, the thin sheet with two circular holes shown in Figure \ref{fig:geometry} occupies the region 
\begin{equation}
\Omega=\left\{-\tau_\mathrm{in}\leqslant\tau\leqslant\tau_\mathrm{in},\quad0\leqslant\xi\leqslant2\pi\right\},
\end{equation} 
where $\tau_\mathrm{in}$ is a positive constant related to the size of the circular hole(s). It then follows from \eqref{eq:circle-x} that 
\begin{equation}
R_\mathrm{in}=\frac{a}{\sinh\tau_\mathrm{in}},\quad \quad O_1=(a\coth\tau_\mathrm{in},0),\quad \quad O_2=(-a\coth\tau_\mathrm{in},0).
\label{eq:radius}
\end{equation}
In the above formulae we have used the Cartesian coordinates to determine the locations of $O_1$ and $O_2$. Moreover, the closest distance between the two holes is 
\begin{equation}
    d=2a\left(\coth\tau_\mathrm{in}-\frac{1}{\sinh\tau_\mathrm{in}}\right).
    \label{eq:distance}
\end{equation}
The scale factors for bipolar coordinates read
\begin{equation}
h_\tau=h_\xi\equiv h=\frac{a}{\cosh\tau-\cos\xi}.
\label{eq:scalor}
\end{equation}

Further details of the bipolar coordinate system are given in \ref{appendix:bipolar}. In this main text, we focus on the broader picture. In the absence of body force, the equilibrium equation for the Cauchy stress tensor $\bm\sigma$ is
\begin{equation}
\div\bm\sigma=\bm 0.
\label{eqn:EqmGeneral}
\end{equation} In addition, the compatibility equation for planar systems provides an extra condition:
\begin{equation}
\Delta\left(\tr\bm\sigma\right)=0,\label{eq:compatibility}
\end{equation}
where `$\Delta$' is the Laplacian operator and `$\tr$' the trace. The boundary conditions are 
\begin{equation}
\begin{aligned}
&\sigma_{\tau\tau}=T_\mathrm{in},\quad\sigma_{\tau\xi}=0,\quad\mathrm{at}~\tau=\pm\tau_\mathrm{in},\\
&\sigma_{\tau\tau}=\sigma_{\xi\xi}=T_\mathrm{out},\quad\mathrm{as}~\tau\rightarrow0,~\xi\rightarrow0.
\end{aligned}\label{eq:BC}
\end{equation}

\section{Exact solution of the elasticity problem}\label{Exact solution}
In this section, we derive the exact solution to the elasticity problem described in Section \ref{Problem formulation}. To this end, we introduce an Airy stress function $\psi(\tau,\xi)$ to ensure that the in-plane equilibrium equation \eqref{eqn:EqmGeneral} is satisfied automatically.  The function $\psi$ is determined by 
  the compatibility equation \eqref{eq:compatibility}, which becomes  a bi-harmonic equation for $\psi$:
\begin{equation}
    \Delta^2\psi=0,
\end{equation}
or, explicitly:
\begin{equation}
    \left(\frac{\partial^4}{\partial\tau^4}+2\frac{\partial^4}{\partial\tau^2\partial\xi^2}+\frac{\partial^4}{\partial\xi^4}-2\frac{\partial^2}{\partial\tau^2}+2\frac{\partial^2}{\partial\xi^2}+1\right)\left(\frac{\psi}{h}\right)=0.
    \label{eq:biharmonic}
\end{equation}

First, we find a particular solution that satisfies the tension condition at infinity of the form \citep{ling1948stresses}:
\begin{equation}
    \psi_0=\frac{1}{2}T_\mathrm{out}(x^2+y^2)=\frac{a}{2}T_\mathrm{out}h\left(\cosh\tau+\cos\xi\right),
\end{equation}
where we have used  \eqref{eq:coordinates} for the second equality. This stress function satisfies equation \eqref{eq:biharmonic}. Specifically, it gives the following uniform stress field in the thin film 
\begin{equation}
T_\mathrm{out}\left(\e_\tau\otimes\e_\tau+\e_\xi\otimes\e_\xi\right).
\end{equation}

Next, we consider the general solution of \eqref{eq:biharmonic}. Since both the geometry and the external loading are symmetric with respect to $x$ (or equivalently $\xi$), we seek a separable solution $\psi_1$ of the form
\begin{equation}
    \frac{\psi_1}{h}=\sum_{n=0}^{\infty}\phi_n(\tau)\cos n\xi.\label{eq:variablese}
\end{equation}
Under this ansatz, equation \eqref{eq:biharmonic} reduces to
\begin{equation}
    \left(\frac{\partial^4}{\partial\tau^4}-2(n^2+1)\frac{\partial^2}{\partial\tau^2}+(n^2-1)^2\right)\phi_n(\tau)=0.
    \label{eq:fourth-ODE}
\end{equation}
The general solutions to \eqref{eq:fourth-ODE} for $n\geqslant2$ are given by
\begin{equation}
\phi_n(\tau)=a_n\cosh(n+1)\tau+b_n\sinh(n+1)\tau+c_n\cosh(n-1)\tau+d_n\sinh(n-1)\tau,
   \label{eq:eigenfunction}
\end{equation}
with $a_n$, $b_n$, $c_n$, and $d_n$ being constants to be determined. For $n=0$ and $n=1$, the corresponding solutions take the form
\begin{align}
    \phi_0(\tau)&=a_0\cosh\tau+b_0\sinh\tau+c_0\tau\cosh\tau+d_0\tau\sinh\tau,\label{eq:phi0}\\
    \phi_1(\tau)&=a_1\cosh2\tau+b_1\sinh2\tau+c_1+d_1\tau.\label{eq:phi1}
\end{align}

We further note that the problem is symmetric in $\tau$, which requires that all coefficients $b_i$ and $d_i$ $(i=1,2,\ldots)$ vanish. Moreover,  the mode $\phi_0$ is discarded because it leads to multi-valued displacements (see \citet{jeffery1921ix} and \citet{ling1948stresses}). Consequently, the eigenfunctions employed in the solution procedure can be expressed in the unified form
\begin{align}
     \phi_n(\tau)&=a_n\cosh(n+1)\tau+c_n\cosh(n-1)\tau,\quad n\geqslant 1.
     \label{eq:phin}
\end{align}
Based on the discussions in \citet{jeffery1921ix}, we  supplement \eqref{eq:variablese} with an additional term to ensure that the solution is globally correct and that the boundary conditions can be satisfied:
\begin{equation}
    K\left(\cosh\tau-\cos\xi\right)\log\left(\cosh\tau-\cos\xi\right),
\end{equation}
where $K$ is a constant. Then, the general solution takes the form
\begin{align}
\notag\frac{\psi}{h}=&~\frac{\psi_0}{h}+\frac{\psi_1}{h}\\=&~
    \frac{a}{2}T_\mathrm{out}\left(\cosh\tau+\cos\xi\right)+K\left(\cosh\tau-\cos\xi\right)\ln\left(\cosh\tau-\cos\xi\right)+\sum_{n=1}^\infty\phi_n(\tau)\cos n\xi.\label{eq:general-sol}
\end{align}
It follows from \eqref{eq:sigmatautau}–\eqref{eq:sigmatauxi} that, once the condition $\psi_1(0,0)=0$ is imposed, all stress components derived from $\psi_1$ vanish at infinity. Hence, the stresses at infinity are solely due to the contribution from $\phi_0$, ensuring that the uniform tension at infinity is preserved. This leads to the condition
\begin{equation}
   \sum_{n=1}^\infty(a_n+c_n)=0.\label{eq:zero-stress}
\end{equation}

Substituting \eqref{eq:general-sol} into \eqref{eq:sigmatautau}-\eqref{eq:sigmatauxi} yields 
\begin{align}
  \notag a\sigma_{\tau\tau}&=aT_\mathrm{out}+\phi_1(\tau)-\frac{1}{2}K\left(\cosh2\tau-2\cosh\tau\cos\xi+\cos2\xi\right)\\&+\frac{1}{2}\sum_{n=1}^\infty\Big((n-1)(n-2)\phi_{n-1}(\tau)-2(n^2-1)\cosh\tau\phi_n(\tau)\notag\\&+(n+1)(n+2)\phi_{n+1}(\tau)-2\sinh\tau\phi_n'(\tau)\Big)\cos n\xi,\label{eq:sigmatautau1}\\\notag  a\sigma_{\xi\xi}&=aT_\mathrm{out}+\phi_1(\tau)-\frac{1}{2}\phi_1''(\tau)+\frac{1}{2}K\left(\cosh2\tau-2\cosh\tau\cos\xi+\cos2\xi\right)\\&\notag+\frac{1}{2}\sum_{n=1}^\infty\Big(2\cosh\tau\phi_n''(\tau)-2\sinh\tau\phi'_n(\tau)+(n+2)\phi_{n+1}(\tau)-\phi_{n+1}''(\tau)-\phi_{n-1}''(\tau)\\&-(n-2)\phi_{n-1}(\tau)\Big)\cos n\xi,\label{eq:sigmaxixi1}\\\notag  a\sigma_{\tau\xi}&=-K\sinh\tau\sin\xi+\frac{1}{2}\sum_{n=1}^\infty\Big(2n\cosh\tau\phi_n'(\tau)-(n+1)\phi_{n+1}'(\tau)\\&-(n-1)\phi_{n-1}'(\tau)\Big)\sin n\xi,\label{eq:sigmatauxi1}
\end{align}
where primes here and hereafter signify differentiation with respect to the implied variable. For example, $\phi_1'(\tau)=\mathrm{d}\phi_1/\mathrm{d}\tau$.

Due to the symmetry of the problem with respect to $\tau$ and  with the aid of \eqref{eq:sigmatautau} and \eqref{eq:sigmatauxi}, the boundary conditions $\eqref{eq:BC}_{1,2}$ can be rewritten in the form
\begin{align}
   \notag &aT_\mathrm{out}+\phi_1(\tau_\mathrm{in})-\frac{1}{2}K\cosh2\tau_\mathrm{in}+\frac{1}{2}\sum_{n=1}^\infty\Big(2K\delta_{1n}\cosh\tau-K\delta_{2n}+(n-1)(n-2)\phi_{n-1}(\tau_\mathrm{in})\\&-2(n^2-1)\cosh\tau\phi_n(\tau_\mathrm{in})+(n+1)(n+2)\phi_{n+1}(\tau_\mathrm{in})-2\sinh\tau_\mathrm{in}\phi_n'(\tau_\mathrm{in})\Big)\cos n\xi=aT_\mathrm{in},\label{eq:normal-bc}\\&
   \frac{1}{2}\sum_{n=1}^\infty\Big(2n\cosh\tau_\mathrm{in}\phi_n'(\tau_\mathrm{in})-K \delta_{1n}\sinh\tau_\mathrm{in}-(n+1)\phi_{n+1}'(\tau_\mathrm{in})-(n-1)\phi_{n-1}'(\tau_\mathrm{in})\Big)\sin n\xi=0.\label{eq:shear-bc}
\end{align}
These two equations give rise to two infinite sets of algebraic relations by requiring that the coefficients of $\cos n\xi$ $(n\geqslant 0)$ and $\sin n\xi$ $(n\geqslant 1)$ vanish independently. In particular, the constant term in \eqref{eq:normal-bc} yields
\begin{align}
    aT_\mathrm{out}-\frac{1}{2}K\cosh2\tau_\mathrm{in}+\phi_1(\tau_\mathrm{in})=aT_\mathrm{in},\label{eq:first-normal}
\end{align}
while the coefficient of $\sin\xi$ in \eqref{eq:shear-bc} reads
\begin{equation}
    \cosh\tau_\mathrm{in}\phi_1'(\tau_\mathrm{in})-\phi_2'(\tau_\mathrm{in})-K\sinh\tau_\mathrm{in}=0.\label{eq:first-shear}
\end{equation}

It is observed that equation~\eqref{eq:first-normal} depends only on $a_1$ and $c_1$ through $\phi_1$. In contrast, equation~\eqref{eq:first-shear} involves $\phi_2$. A careful inspection of \eqref{eq:shear-bc} reveals that, by multiplying the coefficient of each $\sin n\xi$ term by $\mathrm{e}^{-n\tau_\mathrm{in}}$ and summing over all $n$, the contributions from $\phi_n$ with $n\geqslant 2$ can be eliminated. When combined with \eqref{eq:first-normal}, this procedure yields a closed system of two equations for the unknowns $a_1$ and $c_1$, namely
\begin{equation}
\begin{aligned}
    &2\phi_1(\tau_\mathrm{in})=2a(T_\mathrm{in}-T_\mathrm{out})+K\cosh2\tau_\mathrm{in},\\&
    \phi_1'(\tau_\mathrm{in})=2K\mathrm{e}^{-\tau_\mathrm{in}}\sinh\tau_\mathrm{in}.
\end{aligned}
\label{eq:sol-first}
\end{equation}

Moreover, by equating the coefficients of $\cos n\xi$ $(n \geqslant 1)$ in \eqref{eq:normal-bc} and $\sin n\xi$ $(n \geqslant 2)$ in \eqref{eq:shear-bc}, we are able to derive the following recursion relations for $n\geqslant2$ as follows
\begin{equation}
\begin{aligned}
    &
    (n-1)n(n+1)\phi_n(\tau_\mathrm{in})\sinh\tau_\mathrm{in}=\phi_1'(\tau_\mathrm{in})\left(n\cosh n\tau_\mathrm{in}-\coth\tau_\mathrm{in}\sinh n\tau_\mathrm{in}\right)\\&\hspace{54mm}+K\left(\left(n+1\right)\sinh(n-2)\tau_\mathrm{in}-(n-1)\sinh n\tau_\mathrm{in}\right),\\
    &n\phi_n'(\tau_\mathrm{in})\sinh\tau_\mathrm{in}=\phi_1'\sinh n\tau_\mathrm{in}-2K\sinh(n-1)\tau_\mathrm{in}\sinh\tau_\mathrm{in}.
\end{aligned}
\label{eq:recursion}
\end{equation}
(We note that these relations were already identified by \citet{ling1948stresses}, who considered the problem with no internal tension --- i.e.~$T_{\rm in}=0$ in our notation.) Making use of $\eqref{eq:sol-first}_{2}$, straightforward but technical manipulations of \eqref{eq:recursion}, together with standard identities for hyperbolic functions, yield for $n \geqslant 2$
\begin{equation}
    \begin{aligned}
    &\phi_n(\tau_\mathrm{in})=-2\mathrm{e}^{-n\tau_\mathrm{in}}\frac{K(n\sinh\tau_\mathrm{in}+\cosh\tau_\mathrm{in})}{(n-1)n(n+1)},\\&
    \phi_n'(\tau_\mathrm{in})=2\mathrm{e}^{-n\tau_\mathrm{in}}\frac{K\sinh\tau_\mathrm{in}}{n}.
    \end{aligned}\label{eq:sol-n-new}
\end{equation}

We then substitute equation \eqref{eq:phin} into \eqref{eq:sol-first} and \eqref{eq:sol-n-new} to obtain
\begin{equation}
    \begin{aligned}
        a_n&=\dfrac{2K\left(\mathrm{e}^{-n\tau_\mathrm{in}}\sinh n\tau_\mathrm{in}+n\mathrm{e}^{-\tau_\mathrm{in}}\sinh\tau_\mathrm{in}\right)}{n(n+1)\left(\sinh2n\tau_\mathrm{in}+n\sinh2\tau_\mathrm{in}\right)},\quad n\geqslant 1,\\
        c_n&=-\dfrac{2K\left(\mathrm{e}^{-n\tau_\mathrm{in}}\sinh n\tau_\mathrm{in}+n\mathrm{e}^{\tau_\mathrm{in}}\sinh\tau_\mathrm{in}\right)}{n(n-1)\left(\sinh2n\tau_\mathrm{in}+n\sinh2\tau_\mathrm{in}\right)},\quad n\geqslant2,\\
        c_1&=\frac{1}{2}\left(K\tanh\tau_\mathrm{in}\cosh2\tau_\mathrm{in}+2a(T_\mathrm{in}-T_\mathrm{out})\right).
    \end{aligned}\label{eq:sol-coef-sym}
\end{equation}
It follows from \eqref{eq:zero-stress} that
\begin{equation}
    K\left(\frac{1}{2}+\tanh\tau_\mathrm{in}\sinh^2\tau_\mathrm{in}-4\sum_{n=2}^\infty
    M_n\right)=a(T_\mathrm{out}-T_\mathrm{in}),
    \label{eq:k-formula}
\end{equation}
where
\begin{equation}
    M_n=\dfrac{\mathrm{e}^{-n\tau_\mathrm{in}}\sinh n\tau_\mathrm{in}+n\sinh\tau_\mathrm{in}(n\sinh\tau_\mathrm{in}+\cosh\tau_\mathrm{in})}{n(n^2-1)\left(\sinh2n\tau_\mathrm{in}+n\sinh2\tau_\mathrm{in}\right)}.
    \label{eq:Mn}
\end{equation}
In the case $T_\mathrm{in}=0$, equation \eqref{eq:k-formula} recovers the solution given by \citet{ling1948stresses}.

We next briefly show that the infinite series in \eqref{eq:k-formula} is convergent, so that it defines a well-posed expression for $K$. For large $n$, we observe that
\begin{equation}
\sinh n\tau_\mathrm{in}\sim\frac{1}{2}\mathrm{e}^{n\tau_\mathrm{in}},\qquad
\sinh 2n\tau_\mathrm{in}\sim\frac{1}{2}\mathrm{e}^{2n\tau_\mathrm{in}}.
\end{equation}
It then follows that
\begin{equation}
M_n\sim 2\sinh^2\tau_\mathrm{in}\frac{\mathrm{e}^{-2n\tau_\mathrm{in}}}{n},
\end{equation}
and hence
\begin{equation}
|M_n| \leqslant C_0\frac{\mathrm{e}^{-2n\tau_\mathrm{in}}}{n},
\end{equation}
where $C_0$ is a positive constant depending only on $\tau_\mathrm{in}$. Clearly,
\begin{equation}
\sum_{n=2}^{\infty}\frac{\mathrm{e}^{-2n\tau_\mathrm{in}}}{n}
\end{equation}
is convergent. Therefore, the series appearing in \eqref{eq:k-formula} converges for any given $\tau_\mathrm{in}$, ensuring that $K$ is well defined.

 We  obtained an analytical solution given by \eqref{eq:phin} and \eqref{eq:general-sol} with all coefficients determined from \eqref{eq:sol-coef-sym} and \eqref{eq:k-formula}. We observe from \eqref{eq:k-formula} that $K$ is linear in $a$, and hence all coefficients in \eqref{eq:sol-coef-sym} are also linearly proportional to $a$. As a result, the stress components in \eqref{eq:sigmatautau1}–\eqref{eq:sigmatauxi1} are independent of $a$. This is expected, since the two foci $(-a,0)$ and $(a,0)$ can be chosen arbitrarily and their positions should not affect the physical features of the problem. In the next section, we  use this analytical solution to discuss possible wrinkling states.

\section{Onset of wrinkling}\label{onset}

Without loss of generality, in this section we use $a=1$. To facilitate further analysis, we define the tension ratio
\begin{equation}
    \gamma=\frac{T_\mathrm{in}}{T_\mathrm{out}},
\end{equation}
and denote by $\gammacr$ the critical tension ratio at which wrinkles first appear. For the classic Lam\'{e} problem involving a single circular hole, it is well known that, in the thin sheet limit (when the bending stiffness tends to zero), this critical threshold for circumferential wrinkling is $\gammacr=2$ \citep{coman2006localized,coman2007wrinkling,davidovitch2011prototypical}.  We now solve the same problem for two holes, i.e.~we determine the critical tension ratio as $d/a$ varies, $\gammacr(d/a)$.

To measure the relative distance between holes, we introduce the parameter $\zeta=d/R_\mathrm{in}$. Using \eqref{eq:distance}, this parameter can be written as
\begin{equation}
    \zeta=2(\cosh\tau_\mathrm{in}-1).
    \label{eq:zeta}
\end{equation}
Conversely, $\tau_\mathrm{in}$ can be expressed in terms of $\zeta$ as
\begin{equation}
\tau_\mathrm{in}=\operatorname{arccosh}\left(\frac{\zeta}{2}+1\right).
    \label{eq:tauin}
\end{equation}
We note that in the limit $\zeta \to \infty$  the two holes are far apart and the bipolar Lam\'{e} problem reduces to the classical Lam\'{e} problem. We therefore anticipate that $\gammacr(\zeta)\to2$ as $\zeta\to\infty$.

 To determine the value of $\gammacr(\zeta)$, we note that thin sheets with negligible bending stiffness have negligible resistance to compression. As a result, the onset of  wrinkling can be identified by finding when compressive stresses first arise. To this end, we determine the lower principal stress $\sigma_2$ as
\begin{equation}
\sigma_2(\tau,\xi)=\frac{\sigma_{\tau\tau}+\sigma_{\xi\xi}}{2}-\sqrt{\left(\frac{\sigma_{\tau\tau}-\sigma_{\xi\xi}}{2}\right)^2+\sigma^2_{\tau\xi}}.
\label{eq:min-stress}
\end{equation}

\begin{figure}[htbp]
    \centering
\begin{picture}(0,0)(0,0)
\put(156,-4){\Large{$x$}}
\put(-5,90){\Large{$y$}}
\end{picture}
\includegraphics[width=0.7\linewidth]{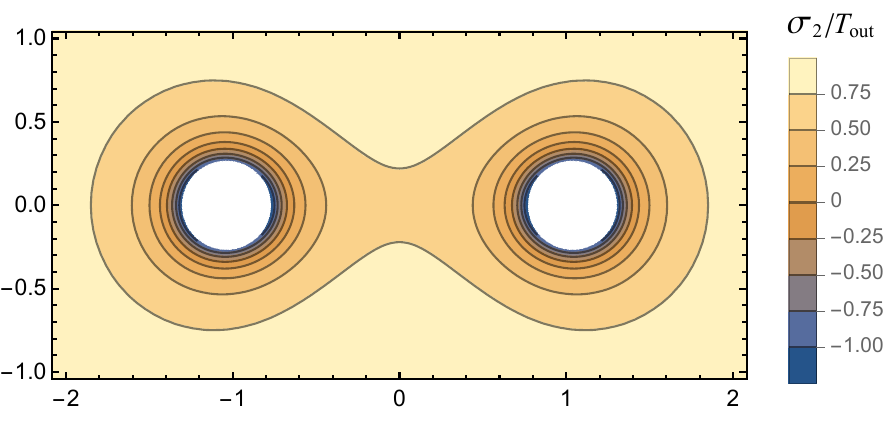}
    \caption{The distribution of the normalized principal stress $\sigma_2/T_\mathrm{out}$ for $\tau_\mathrm{in}=2$ and $\gamma=3$.}
    \label{fig:stress-distribution}
\end{figure}
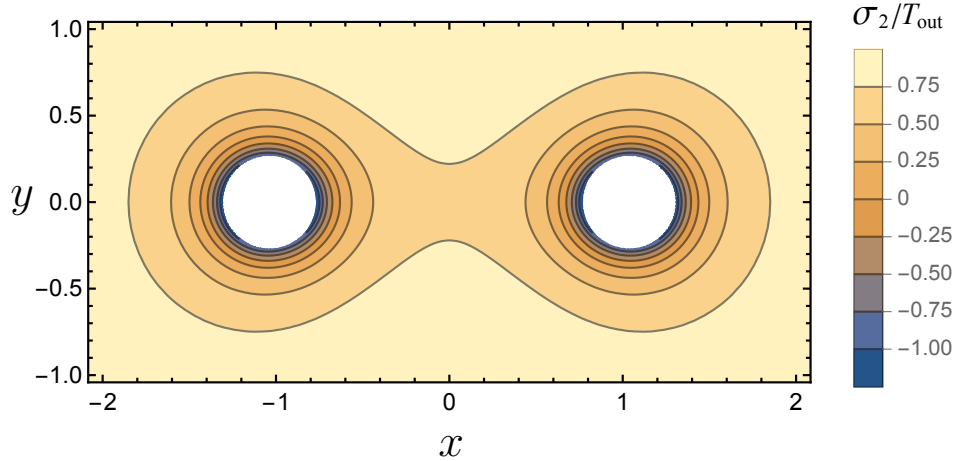

Figure~\ref{fig:stress-distribution} illustrates the spatial distribution of the normalized stress component $\sigma_2/T_\mathrm{out}$ for a tension ratio $\gamma = 3$ and $\tau_\mathrm{in} = 2$. The two circular holes are represented by the white disks, while the black curves indicate contours of constant stress. For this configuration, the radius of each circular hole is $R_\mathrm{in} = 1/\sinh 2 \approx 0.275721$, and the distance between the two holes is $d \approx 1.52319$ (see \eqref{eq:distance}). It is evident from the figure that regions of compressive stress develop in the vicinity of both holes. Moreover, the contours corresponding to a given stress level exhibit shapes resembling Cassini ovals. We emphasize that the analytical solution in the previous section is valid only for the primary deformation, for which no instability is present --- since very thin sheets cannot support compression, the compressive stress observed in Figure \ref{fig:stress-distribution} would not be observed experimentally.

\begin{figure}[h]
    \centering
    \subfigure{\includegraphics[width=0.45\linewidth]{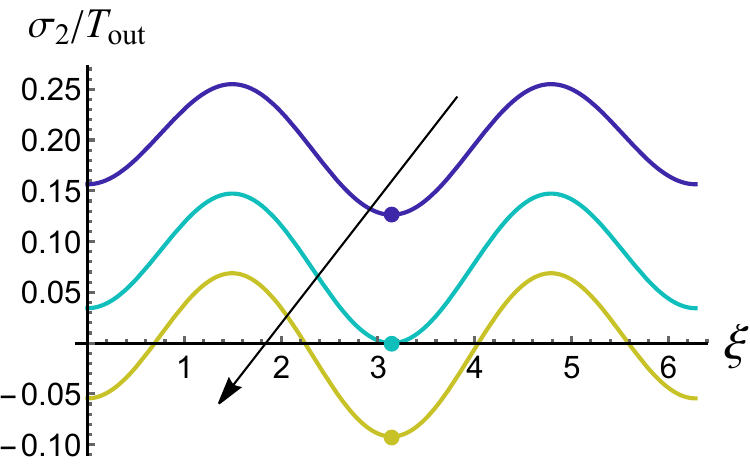}}\quad
    \subfigure{\includegraphics[width=0.45\linewidth]{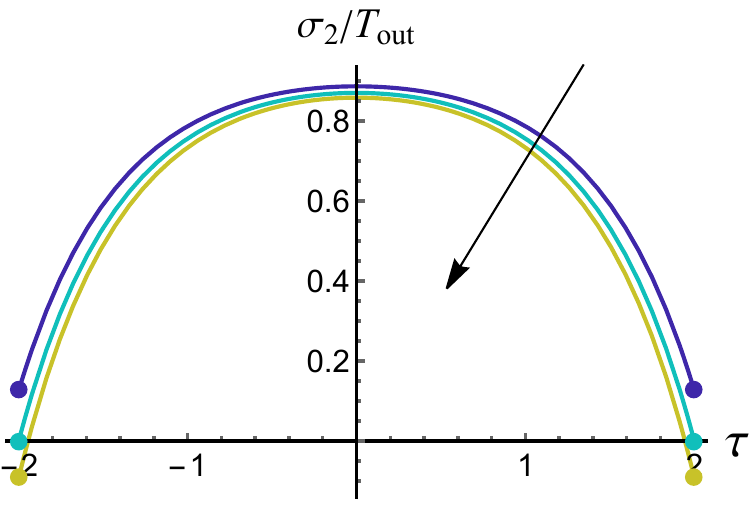}}
    \caption{Dependence of the normalized stress $\sigma_2/T_\mathrm{out}$ on $\xi$ (left panel) and $\tau$ (right panel) for $\tau_\mathrm{in}=2$ and $\gamma\in\{1.8,\,\gamma_\mathrm{cr}=1.9158,\,2\}$. The arrows in each figure indicate the direction of increasing $\gamma$. In the left panel, $\tau=2$, while in the right panel $\xi=\pi$. The global minimum is highlighted with points in each panel.}
    \label{fig:sigma-min}
\end{figure}

To identify where the minimum value of $\sigma_2$ occurs in the sheet, we plot in Figure~\ref{fig:sigma-min} the normalized stress as either $\tau$ or $\xi$ varies. From the left panel, we observe that $\sigma_2$ attains its global minimum at $\xi=\pi$. By contrast, the right panel shows that $\sigma_2$ reaches its minimum at $\tau=\pm\tau_\mathrm{in}$. Accordingly, the lowest stress in the sheet is attained at $(\tau,\xi)=(\pm\tau_\mathrm{in},\pi)$, and the critical tension ratio $\gamma_\mathrm{cr}$ for the onset of wrinkling can be identified from the condition
\begin{equation}
    \sigma_2(\tau_\mathrm{in},\pi,\gamma_\mathrm{cr})=0.
    \label{eq:onset}
\end{equation}

\begin{figure}[htbp]
    \centering
    \includegraphics[width=0.8\linewidth]{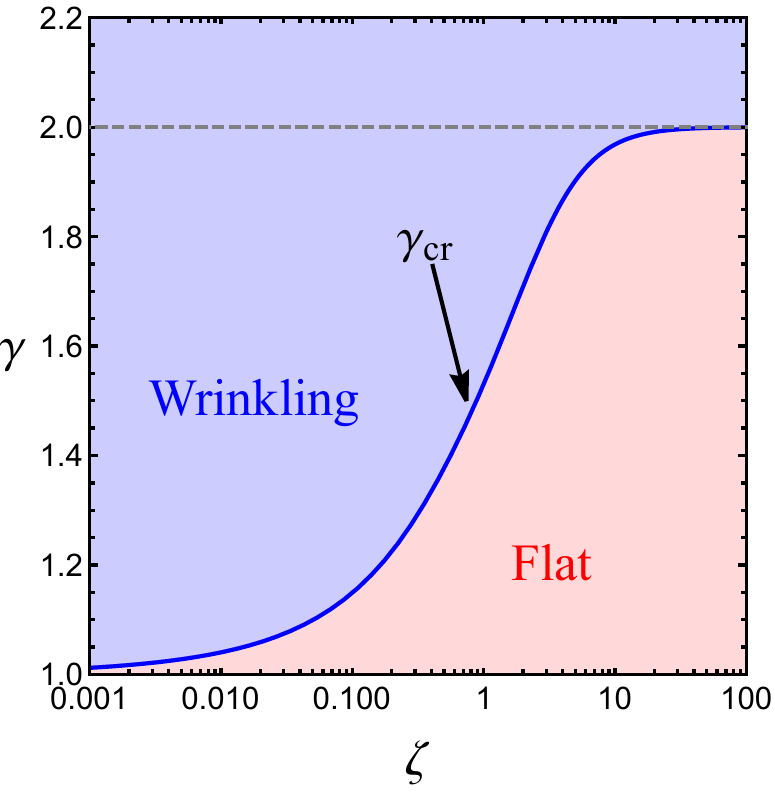}
    \caption{Phase diagram showing the transition between wrinkled and flat configurations in the $\gamma$–$\zeta$ plane. The blue solid line denotes the critical tension ratio $\gamma_\mathrm{cr}$, while the gray dashed line (corresponding to $\gamma=2$) indicates the critical tension ratio for the Lam\'{e} problem.}
    \label{fig:gamma-zeta}
\end{figure}

Substituting \eqref{eq:tauin} into the critical condition \eqref{eq:onset} yields an expression for the critical tension ratio $\gamma_\mathrm{cr}$ for a given $\zeta$. The resulting behavior of $\gammacr(\zeta)$ and a  phase diagram are shown in Figure~\ref{fig:gamma-zeta}; this illustrates the regions in parameter space associated with wrinkled or flat sheets. The blue curve gives the threshold $\gammacr$. As $\gamma$ increases from one, for a given $\zeta$, wrinkling appears for $\gamma>\gamma_\mathrm{cr}$. The function $\gamma_\mathrm{cr}(\zeta)$ is a monotonically increasing function that tends asymptotically to $\gamma=2$ as  $\zeta \to \infty$. We note also that the limit  $\zeta \to 0$, corresponds to the limit in which the two circular holes just touch. In this case, the critical tension ratio approaches $1$. We discuss this limit in more detail later, but for now note that, in comparison with the single-hole case, the presence of an additional hole always promotes an earlier onset of instability. A similar early wrinkling phenomenon was reported by \citet{andrade2019pre} when the circular hole in the classical Lam\'{e} problem is replaced by an elliptical hole. We now turn to determining asymptotic relationships for $\gammacr$ that are valid when $\zeta$ is large or small.

\subsection{Large hole separations: $\zeta\gg1$}
In the large separation limit, we consider an asymptotic expansion in $1/\zeta$:
\begin{equation}
    \frac{\sigma_2(\zeta,\pi,\gamma_\mathrm{in})}{T_\mathrm{out}}=2-\gamma+\frac{4(1-\gamma)}{\zeta^2}+\frac{8(\gamma-1)}{\zeta^3}+\frac{8(1-\gamma)}{\zeta^4}+O\left(\zeta^{-5}\right).
\end{equation}
Substituting this expansion into \eqref{eq:onset}, we obtain a four-term asymptotic approximation for the critical tension ratio,
\begin{equation}
    \gamma_\mathrm{cr}=2-\frac{4}{\zeta^2}+\frac{8}{\zeta^3}+\frac{8}{\zeta^4}+O\left(\zeta^{-5}\right).
    \label{eq:asy-gamma-four}
\end{equation}
Based on this result, we further apply the Pad\'{e} approximant to obtain a rational function \citep{Baker_Graves-Morris_1996}
\begin{equation}
    \gamma_\mathrm{cr}\approx\frac{2\zeta^2+4\zeta+8}{\zeta^2+2\zeta+6}.
    \label{eq:pade}
\end{equation}
The expansion of \eqref{eq:pade} in powers of $1/\zeta$ coincides with \eqref{eq:asy-gamma-four} up to the truncated order, while providing a substantially improved approximation and thereby extending the effective range of validity of the asymptotic analysis.

\begin{figure}[h]
    \centering
   \includegraphics[width=0.5\linewidth]{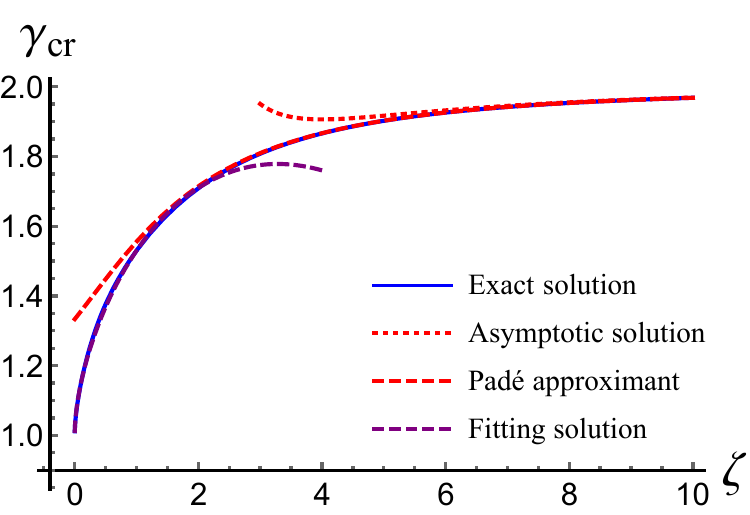}
    \caption{Comparisons between the exact solution (blue solid curve) and the approximate solutions. The asymptotic prediction \eqref{eq:asy-gamma-four} (red dotted) and its Pad\'e approximant \eqref{eq:pade} (red dashed) recover the numerically-determined exact behavior with the latter being essentially exact for $\zeta>2$. Moreover, the estimate \eqref{eq:fitting} (purple dashed) agrees well with the exact solution for $\zeta<2.5$.}
    \label{fig:comparison}
\end{figure}

The two approximate solutions valid when $\zeta\gg1$, i.e.~\eqref{eq:asy-gamma-four} and \eqref{eq:pade}, are compared with the (numerically-determined) exact solution in Figure~\ref{fig:comparison}. It is observed that the four-term asymptotic solution \eqref{eq:asy-gamma-four} (red dotted curve) remains accurate up to $\zeta \sim 6$, which is remarkable given that the underlying assumption of the asymptotic analysis is $\zeta \gg 1$. More importantly, the Pad\'{e} approximant \eqref{eq:pade} (red dashed curve) exhibits excellent agreement with the exact solution even down to $\zeta \sim 1.5$. This demonstrates that the Pad\'{e} approximant provides a significantly improved approximation and can be reliably used to predict the wrinkling threshold over a much wider range of $\zeta$.

\subsection{Small hole separations: $\zeta\ll1$}

From Figure~\ref{fig:gamma-zeta}, we also observe that $\gamma_\mathrm{cr}\to 1$ as $\zeta\to 0$. This implies that when the two holes are sufficiently close, wrinkling occurs as soon as $T_\mathrm{in}$ slightly exceeds $T_\mathrm{out}$. In this limit, it can be shown that $\sum_{n=2}^\infty M_n$ takes the limit \citep{ling1948stresses}
\begin{equation}
  \sum_{n=2}^\infty M_n = \sum_{n=2}^\infty \frac{1}{n(n^2-1)}-2\tau_\mathrm{in}^2\int_0^\infty\frac{\sinh^2\varsigma-\varsigma^2}{\varsigma^3(\sinh2\varsigma+2\varsigma)}\mathrm{d}\varsigma.
\end{equation}
In view of this, we deduce that $K$ is well defined as $\zeta \rightarrow 0$ (or equivalently $\tau_\mathrm{in} \rightarrow 0$). However, it is known that the stresses when expressed as an infinite series \eqref{eq:sigmatautau1}-\eqref{eq:sigmatauxi1}, do not converge uniformly when $\zeta\rightarrow0$ \citep{callias1989singular,zimmerman1988stress}. Consequently, the limit operator and the summation operator cannot be interchanged, and we are therefore unable to derive a valid asymptotic solution for the critical tension ratio $\gamma_\mathrm{cr}$. Nevertheless, Figure~\ref{fig:gamma-zeta} suggests that the asymptotic behaviors in the limits $\zeta \to 0$ and $1/\zeta \to 0$ are qualitatively similar. In addition, as detailed in \citet{zimmerman1988stress,callias1993singularity,wu1996singular}, the maximum stress in this case blows up in the order of $\zeta^{1/2}$ if $T_\mathrm{in}$ vanishes. Motivated by this and the ansatz in \eqref{eq:asy-gamma-four}, we fit the exact solution using a series expansion in powers of $\zeta^{1/2}$. Retaining only two significant figures in the fitted coefficients, we obtain 
\begin{equation}
    \gamma_\mathrm{cr}\approx1+\frac{2}{5}\zeta^{1/2}+\frac{27}{100}\zeta-\frac{7}{50}\zeta^{3/2}.
    \label{eq:fitting}
\end{equation}

A comparison between the fitted formula \eqref{eq:fitting} and the exact solution is  presented in Figure~\ref{fig:comparison}. It can be seen that the  expression \eqref{eq:fitting} provides excellent agreement with the exact solution even for values of $\zeta$ as large as $2.5$.

Combining \eqref{eq:pade} and \eqref{eq:fitting}, we provide a complete set of explicit solutions for the wrinkling initiation over the entire range of $\zeta$, which is useful for accurately predicting the onset of wrinkling, as shown in Figure \ref{fig:comparison}.



\section{Extent of wrinkles}\label{extent-of-wrinkles}

\subsection{Near-threshold wrinkling}\label{near-threshold}

As  $\gamma$ exceeds $\gamma_\mathrm{cr}$, wrinkles are induced to release compressive stresses in the sheet. Consequently, the stresses must be updated to account for this change. In general, tension-field theory can be used to approximate the stress distribution by enforcing $\sigma_2=0$ in regions where compressive stress would otherwise occur \citep{Mansfield1969,pipkin1986relaxed,steigmann1989finite,steigmann1989wrinkling,steigmann1990tension}. In this framework, the whole sheet is divided into two regions connected through an interface: an inner wrinkled region where $\sigma_2=0$, and an outer flat tensile region. For the Lam\'{e} problem, \citet{davidovitch2011prototypical} applied tension-field theory to  obtain the wrinkled region. In particular, they showed that the maximum stress decays in inverse proportion to the distance from the hole center. When compared to the faster (inverse square) decay that occurs in the purely tensile case, we see that wrinkling allows the stress signal to spread over longer distances. In particular, while the size of the region in which compression is observed would grow like $\sqrt{\gamma-\gammacr}$ if compressive stress were resisted, wrinkling (and the associated relaxation of compression) means that the size of the wrinkled region grows linearly with $\gamma>\gammacr$.

Although tension-field theory is a powerful tool for determining the actual extent of wrinkling, its direct application to the present problem is not straightforward because of the loss of radial symmetry. In particular, the approach employed by \citet{davidovitch2011prototypical} cannot  readily be extended to this setting. Nevertheless, when the tension ratio $\gamma$ is only slightly above the critical value $\gamma_\mathrm{cr}$, that is, in the near-threshold regime, we expect that the wrinkled region may still be described using the purely tensile elastic solution presented in Section~\ref{Exact solution} \citep{davidovitch2011prototypical}. Motivated by this observation, we focus in this subsection on analyzing the wrinkled region in the near-threshold case, with the aim of providing new insight into  where wrinkles first emerge in an elastic sheet containing two circular holes.

\subsubsection{Numerical results for intermediate hole separations: $\zeta=O\left(1\right)$}

\begin{figure}[h]
    \centering
    \includegraphics[width=0.95\linewidth]{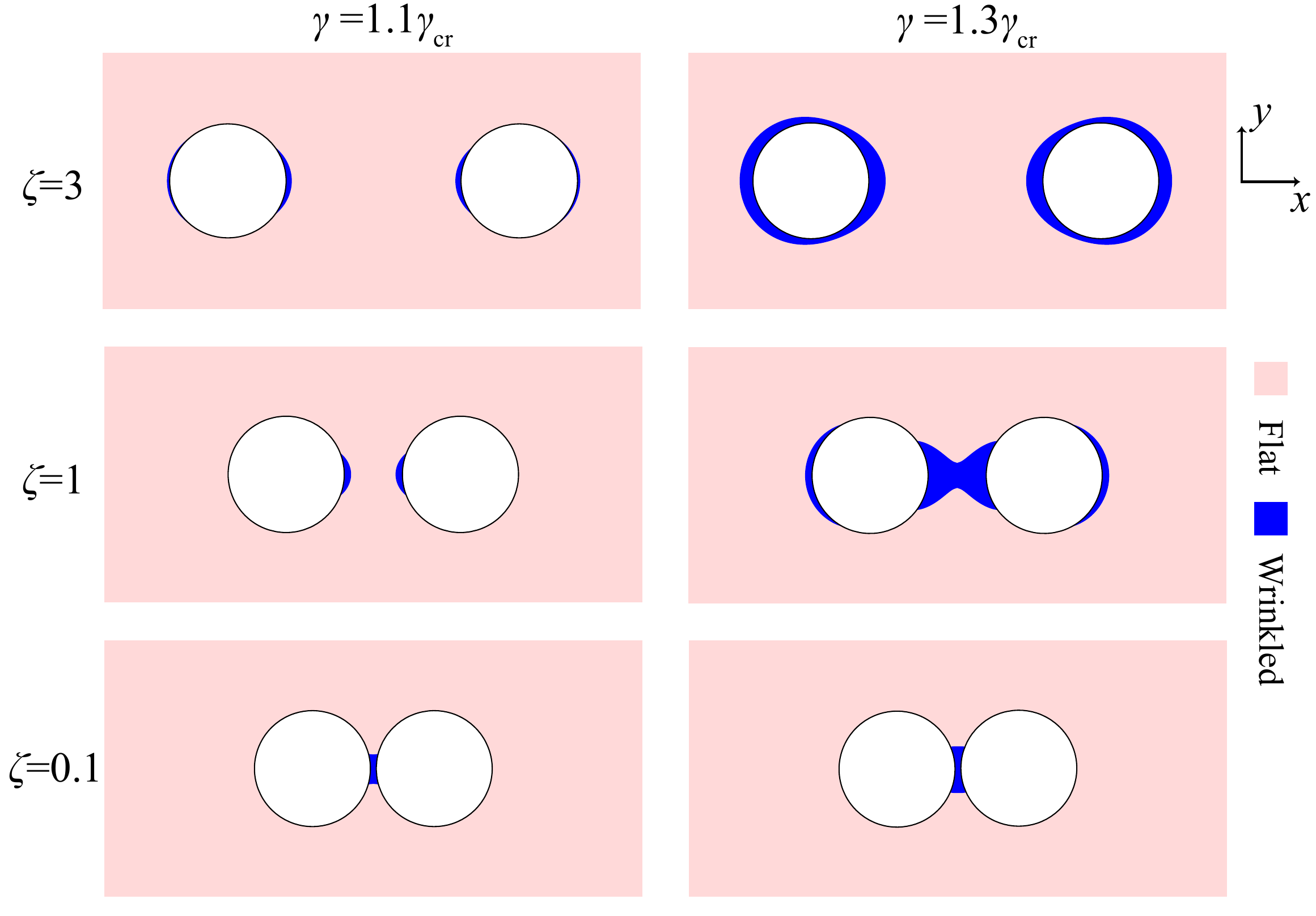}
    \caption{Representation of the wrinkled regions (highlighted in blue) for  $\gamma$ close to $\gamma_\mathrm{cr}$ and different values of the scaled hole separation $\zeta=d/R_{\rm in}$. }
    \label{fig:NT}
\end{figure}

When the two holes are far apart, the system behaves like the classical Lam\'{e} problem. For this reason, we first focus on configurations in which the two holes are relatively close. To examine the near-threshold regime, the tension ratio is prescribed as $\gamma = 1.1\gamma_\mathrm{cr}$ or $\gamma = 1.3\gamma_\mathrm{cr}$, allowing us to qualitatively trace the evolution of the wrinkled region. The boundary of the wrinkled area is identified by locating the zero contour of the lower principal stress $\sigma_2$ for a given $\gamma > \gamma_\mathrm{cr}$. The resulting wrinkled regions are shown in blue in Figure~\ref{fig:NT}. We consider three representative values of the hole separation parameter, namely $\zeta = 0.1,1,$ and $3$. 

For $\zeta = 3$ and $\gamma \gtrsim \gamma_\mathrm{cr}$, wrinkles start at the front and rear of each hole. Recall that wrinkle initiation occurs at $(\pm\tau_\mathrm{in}, \pi)$, corresponding to the Cartesian locations $(-d/2, 0)$ and $(d/2, 0)$, where $d$ denotes the distance between the two holes. Consequently, wrinkles first develop in the forward and backward directions along the $x$-axis. As $\gamma$ increases to $1.3\gamma_\mathrm{cr}$, wrinkles are mainly concentrated on both lateral sides of the holes. This suggests that when $\gamma$ is sufficiently large, the wrinkled regions surrounding the two holes will eventually merge at the center of the sheet.
When the holes are closer ($\zeta = 1$), wrinkles initially appear only on the facing sides of the holes. With further increases in $\gamma$, wrinkles extend into the region between the two holes merging together, while additional wrinkles also develop on the outer sides. This behavior is consistent with the experimental observations reported in Figure 4.4(B) of \citet{huang2010wrinkling}. Although the overall wrinkled extent is significantly amplified due to stress redistribution in the sheet, the present near-threshold analysis is still able to capture the primary geometric features of the wrinkled patterns.
Finally, we consider the extreme case $\zeta = 0.1$, where the two holes are very close. In this regime, wrinkles are confined to the region between the holes and gradually expand in the $y$-direction as $\gamma$ increases from $1.1\gamma_\mathrm{cr}$ to $1.3\gamma_\mathrm{cr}$.

\subsubsection{Widely separated holes: $\zeta\gg1$}

Turning now to the case $\zeta\gg1$ where the two holes are far apart we recall that we were previously able to obtain an asymptotic solution, \eqref{eq:asy-gamma-four}, for the critical tension ratio. In the same way, we can carry out an asymptotic analysis to derive explicit formulae for the stress components when $\zeta\gg1$.  Details are provided in \ref{appendixa}. In summary,  we are able to express the principal stress $\sigma_2$ in \eqref{eq:min-stress} as a series of $\zeta$ (using $\eqref{eq:stress-far}$). An approximation for the boundary of the wrinkled area may then be determined from the condition $\sigma_2=0$. We find that the boundary of the wrinkled region is given by $r_1=W_{\rm NT}(\theta_1)$ where
\begin{equation}
   W_\mathrm{NT}(\theta_1)=(\gamma-1)^{1/2}R_\mathrm{in}+\frac{R_\mathrm{in}\cos\theta_1}{\zeta}+\frac{1}{4\zeta^2}\left(\frac{\left(2\gamma^2-4\gamma+9\right)\cos2\theta_1}{\sqrt{\gamma-1}}-8\cos\theta_1\right)R_\mathrm{in}+O\left(\zeta^{-3}\right),
   \label{eq:wrinkled-boundary}
\end{equation}
with $\theta_1$ as defined in Figure \ref{fig:bipolar}. It is found that \eqref{eq:wrinkled-boundary} reduces to the $(\gamma-1)^{1/2}R_\mathrm{in}$ as $\zeta\rightarrow\infty$, which is identical to the result for the Lam\'{e} problem \citep{davidovitch2011prototypical}.

\begin{figure}[htbp]
    \centering
    \includegraphics[width=0.95\linewidth]{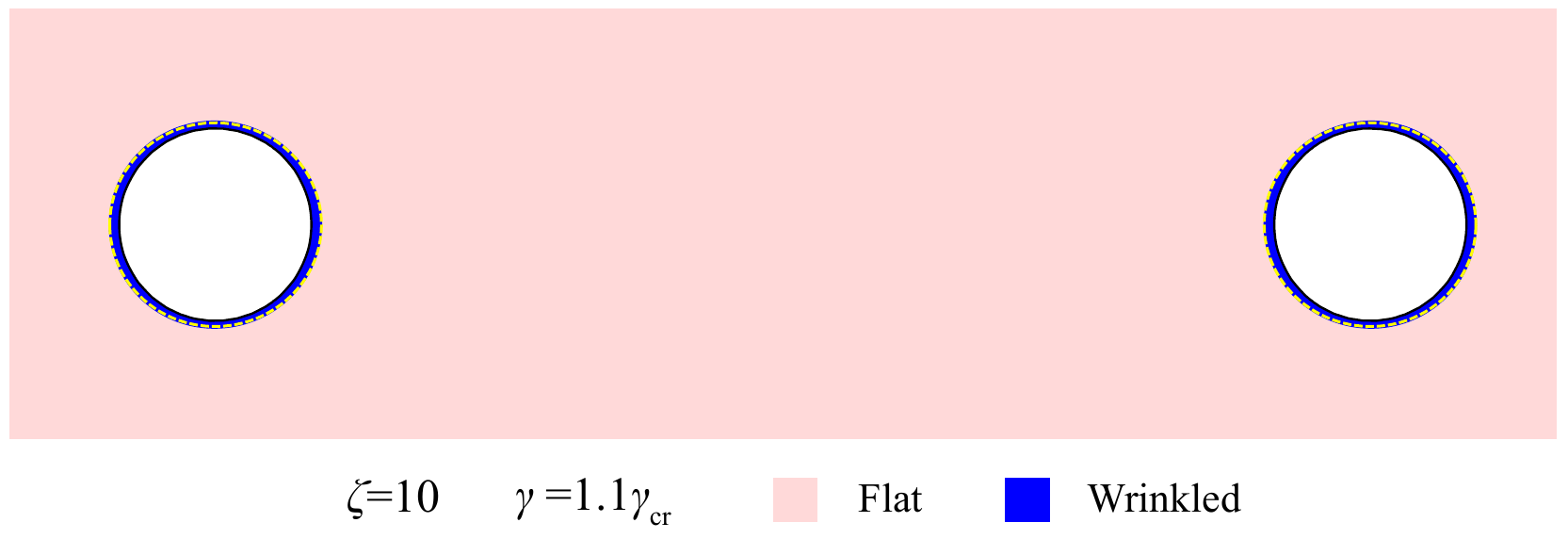}
    \caption{The extent of wrinkles in the near-threshold regime when the two holes are far apart. The  dashed curve indicates the approximate boundary of wrinkled region predicted by \eqref{eq:wrinkled-boundary}, which coincides well with the extent of the region in which compression exists (shaded blue).}
    \label{fig:zeta10}
\end{figure}

The example, shown in Figure~\ref{fig:zeta10} demonstrates 
excellent agreement between the two solutions. When the two holes are well separated, they do not interact with each other and the wrinkling behavior is locally that of the classical Lam\'{e} problem:   wrinkles develop more or less uniformly around the periphery of the hole.

\subsection{Far-from-threshold wrinkling when the two holes are far apart}\label{FFT}

The stress distribution shown in Figure~\ref{fig:stress-distribution} is replaced by an energetically favorable wrinkled state when $\gamma>\gamma_\mathrm{cr}$. The main idea  when applying the tension-field theory to the Lam\'{e} problem  is to set the compressive stress to zero and then solve the modified equilibrium equations \citep{pipkin1986relaxed,steigmann1990tension,davidovitch2011prototypical}. Making use of \eqref{eq:min-stress}, we find that, in the wrinkled region,
\begin{equation}
    \sigma_2=0\Longrightarrow \sigma_{\tau\xi}^2-\sigma_{\tau\tau}\sigma_{\xi\xi}=0\Longleftrightarrow\det\boldsymbol{\sigma}=0.\label{eq:detsigma}
\end{equation}
We note that the final relation is applicable in any coordinate system.

In contrast to the Lam\'{e} problem, the $\tau$- and $\xi$-axes no longer coincide with the principal directions. Consequently, equations  \eqref{eqn:EqmGeneral} and \eqref{eq:detsigma} must be solved simultaneously. Moreover, the profile of the boundary separating the wrinkled and flat regions is typically unknown, which makes the application of tension-field theory challenging in general. Nevertheless, in the far limit of well-separated holes, $\zeta \rightarrow \infty$,  we can perform a perturbation analysis based on the exact far-from-threshold solutions \citep{davidovitch2011prototypical}.

To do so, we  take $1/\zeta$ as a small parameter and use it to construct an asymptotic solution. Without loss of generality, we fix the right hole, and move the left hole away  in Figure~\ref{fig:bipolar}, which is equivalent to restricting attention to the right half-plane. In doing so, we select a polar coordinate system centered at the right center $O_1$. Accordingly, the coordinate $r$ measures the distance from any material point to this new origin, while $\theta$ denotes the usual polar angle. Note that $r=a/\sinh\tau\sim r_1$ (see \eqref{eq:r1}) and $\theta\sim\theta_1$ for large $\zeta$, where $r_1$ and $\theta_1$ are defined in Figure \ref{fig:bipolar}. We further assume that when $\gamma$ is much larger than the critical value $\gamma_\mathrm{cr}$, the  wrinkled region is confined to $r\leqslant W_{\rm FT}(\theta)$ where 
\begin{equation}
    W_\mathrm{FT}(\theta)=R_\mathrm{in}\frac{\gamma}{2}+\frac{W^{(1)}(\theta)}{\zeta }+O\left(\zeta^{-2}\right).
    \label{eq:L}
\end{equation}
Here, the leading-order term is the far-from-threshold solution of the Lam\'{e} problem \citep{davidovitch2011prototypical}, which is axisymmetric.  $W^{(1)}(\theta)$ (and all quantities with superscript $(1)$)  is then an unknown function, to be determined in the subsequent analysis, that captures the leading-order deviations from axisymmetry.

In the wrinkled region, we seek a perturbative solution of the form
\begin{equation}
\sigma_{rr}^\mathrm{in}=T_\mathrm{in}\frac{R_\mathrm{in}}{r}+\frac{\sigma_{rr}^{(1)}(r,\theta)}{\zeta }+O\left(\zeta^{-2}\right),
\quad\sigma_{r\theta}^\mathrm{in}=\frac{\sigma_{r\theta}^{(1)}(r,\theta)}{\zeta }+O\left(\zeta^{-2}\right),\quad
\sigma_{\theta\theta}^\mathrm{in}=\frac{ \sigma_{\theta\theta}^{(1)}(r,\theta)}{\zeta^2}+O\left(\zeta^{-3}\right),\label{eq:sol-in}
\end{equation}
where the leading-order (and axisymmetric) solution is given in \citet{davidovitch2011prototypical}, and the perturbation orders are chosen to obtain the higher-order corrections.

The leading-order terms satisfy exactly the boundary conditions at $r=R_\mathrm{in}$,
\begin{equation}
\sigma_{rr}^\mathrm{in}=T_\mathrm{in},\quad \sigma_{r\theta}^\mathrm{in}=0.
\end{equation}
To obtain a non-trivial solution for the higher-order terms, we relax the boundary conditions and instead require that the resultant forces satisfy the null boundary conditions in an average sense:
\begin{equation}
\int_0^{2\pi} R_\mathrm{in}\sigma_{rr}^{(1)} \mathrm{d}\theta
=\int_0^{2\pi} R_\mathrm{in}\sigma_{r\theta}^{(1)} \mathrm{d}\theta
=0,\quad \mbox{at } r=R_\mathrm{in}.
\label{eq:relaxed-bc}
\end{equation}
From the equilibrium equations in polar coordinates, retaining terms up to $O\left(\zeta^{-2}\right)$, leads to
\begin{equation}
         \frac{\partial\left(r\sigma_{rr}^{(1)}\right)}{\partial r}+\dfrac{\partial\sigma_{r\theta}^{(1)}}{\partial\theta}-\frac{\sigma_{\theta\theta}^{(1)}}{\zeta}=0,\quad
        r\dfrac{\partial\sigma_{r\theta}^{(1)}}{\partial r}+\dfrac{\partial\sigma_{\theta\theta}^{(1)}}{\partial\theta}\frac{1}{\zeta}+ 2\sigma_{r\theta}^{(1)}=0.
\end{equation}
It then follows from these equations that
\begin{equation}
    \sigma_{r\theta}^{(1)}=\dfrac{C_1(\theta)}{r^2},\quad \sigma_{rr}^{(1)}=\frac{C_1'(\theta)}{r^2}+\frac{C_2(\theta)}{r},
    \label{tenfield-sol}
\end{equation}
where $C_1(\theta)$ and $C_2(\theta)$ are unknown functions. The auxiliary equation $\eqref{eq:detsigma}_3$ yields
\begin{equation}
   \left(T_\mathrm{in}\frac{R_\mathrm{in}}{r}+\frac{1}{\zeta} \sigma_{rr}^{(1)}\right)\frac{1}{\zeta^2}\sigma_{\theta\theta}^{(1)}-\frac{1}{\zeta^2}\left(\sigma_{r\theta}^{(1)}\right)^2=0.
   \label{eq:auxiliary}
\end{equation}

We solve the $O\left(\zeta^{-2}\right)$ equation of \eqref{eq:auxiliary} to obtain
\begin{equation}
    \sigma_{\theta\theta}^{(1)}=\frac{C_1^2(\theta)}{r^3T_\mathrm{in}R_\mathrm{in}}.
\end{equation}
Note that the continuity condition at $r=W_{\rm FT}$ imposes 
\begin{equation}
   \bm n\cdot \bm\sigma^\mathrm{in}\bm n=\bm n\cdot \bm\sigma^\mathrm{out}\bm n,\quad  \bm t\cdot \bm\sigma^\mathrm{in}\bm n=\bm t\cdot \bm\sigma^\mathrm{out}\bm n, 
   \label{eq:continuity-far}
\end{equation}
where the superscript `in' (`out') indicates the stress tensor inside  (outside) the wrinkled region and the normal and tangent vectors of the surface $W_{\rm FT}=R_\mathrm{in}\gamma/2+W^{(1)}(\theta)/\zeta$ is given by
\begin{equation}
    \bm{n}=-\bm e_r+2\left(W^{(1)}\right)'/(\zeta R_\mathrm{in}\gamma)\bm e_\theta,\quad  \bm{t}=\bm e_\theta+2\left(W^{(1)}\right)'/(\zeta R_\mathrm{in}\gamma)\bm e_r.
\end{equation}

Next, we consider the solutions in the flat region. Note that in the large-separation limit, the stresses recover their counterparts in the classic Lam\'{e} problem, as given in \eqref{eq:stress-far}. Under this correspondence, we give the outer solution through replacing $R_\mathrm{in}$ by $W_{\rm FT}$ and $T_\mathrm{in}$ by $T(W_{\rm FT})$ in \eqref{eq:stress-far}. Accordingly, we obtain
\begin{equation}
    \begin{aligned}
    &\sigma_{rr}^\mathrm{out}=T_\mathrm{out}+\frac{(T_\mathrm{in}-T_\mathrm{out})\gamma^2R_\mathrm{in}^2}{4r^2}+\left(\frac{\gamma^2R_\mathrm{in}^2(T_\mathrm{in}-T_\mathrm{out})\left(\cos\theta\gamma R_\mathrm{in}-2W^{(1)}(\theta)\right)}{4r^3}\right)\frac{1}{\zeta}+O\left(\zeta^{-2}\right),\\
        &\sigma_{r\theta}^\mathrm{out}=\frac{\gamma^3R_\mathrm{in}^3(T_\mathrm{in}-T_\mathrm{out})}{4r^3\zeta}\sin\theta+O\left(\zeta^{-2}\right),\\
        &\sigma_{\theta\theta}^\mathrm{out}=T_\mathrm{out}-\frac{(T_\mathrm{in}-T_\mathrm{out})\gamma^2R_\mathrm{in}^2}{4r^2}-\left(\frac{\gamma^2R_\mathrm{in}^2(T_\mathrm{in}-T_\mathrm{out})\left(\cos\theta\gamma R_\mathrm{in}-2W^{(1)}(\theta)\right)}{4r^3}\right)\frac{1}{\zeta}+O\left(\zeta^{-2}\right).
    \end{aligned}\label{eq:sol-out}
\end{equation}
Since the hoop stress $\sigma_{\theta\theta}$ is continuous across the boundary between the wrinkled and non-wrinkled regions \citep{davidovitch2011prototypical} we can substitute \eqref{eq:sol-in} and \eqref{eq:sol-out} into \eqref{eq:continuity-far}, to obtain
\begin{equation}
\begin{aligned}
    W^{(1)}(\theta)=\frac{\gamma\cos\theta}{4}R_\mathrm{in},\quad C_1(\theta)=\frac{R_\mathrm{in}^2T_\mathrm{in}}{4}\gamma\sin\theta,\quad C_2=0.
\end{aligned}
\end{equation}
It can be readily verified that the boundary conditions in \eqref{eq:relaxed-bc} are satisfied. Finally, for the far-from-threshold regime, the extent of the wrinkled region is given by ($a$ can be replaced by $\zeta$ using \eqref{eq:a-zeta})
\begin{equation}
    W_{\rm FT}=\frac{\gamma}{2}R_\mathrm{in}+\frac{\gamma\cos\theta}{4\zeta}R_\mathrm{in}+O\left(\zeta^{-2}\right).
\end{equation}
Since the second term depends on $\theta$, the longest wrinkles appear at $\theta=0$ and $\pi$, corresponding to two points on the $x$-axis located at the rear sides of the holes.

\section{Experimental validation}\label{experiments}
\begin{figure}[h]
    \centering
    \includegraphics[width=0.9\linewidth]{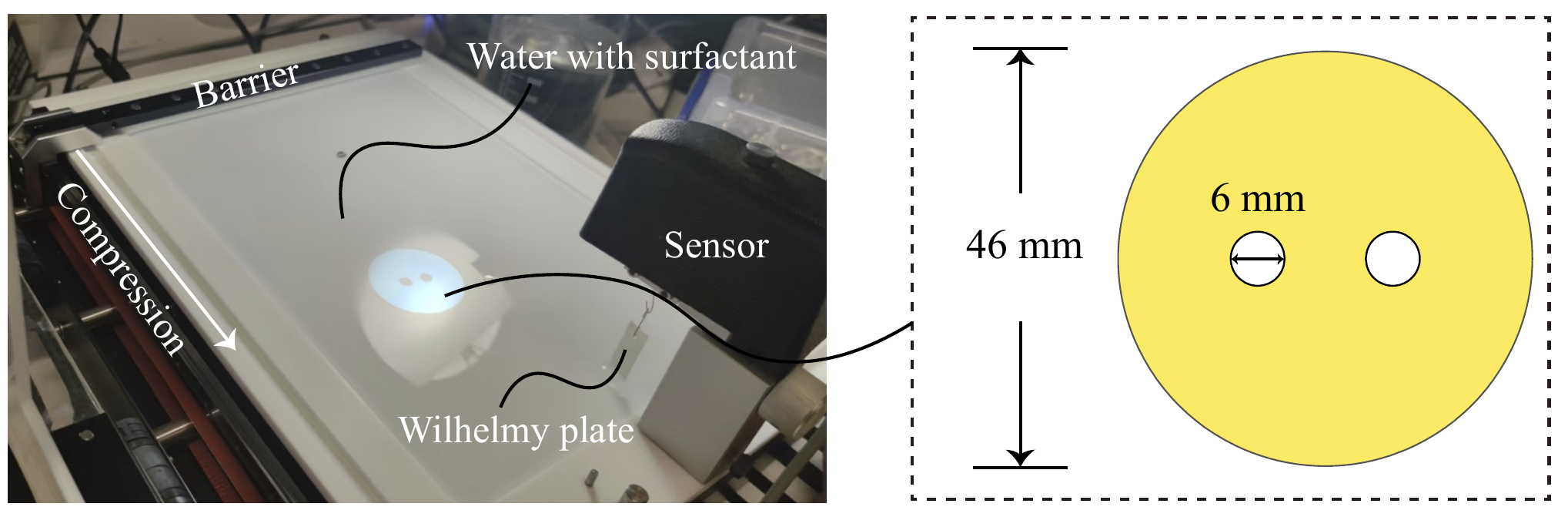}
    \caption{Illustration of the experimental setup. (Left) A Langmuir trough (NIMA 601) is used to load a thin film containing two circular holes by varying the concentration of surfactant at the interface outside the sheet. The interfacial tension of the external interface is measured using a Wilhelmy plate. (Right) Geometry of the sheet: The sheet has an overall circular shape with a diameter of $46~\mathrm{mm}$, while each of the two circular holes has a diameter of $6~\mathrm{mm}$. }
    \label{fig:experiment}
\end{figure}

In this section, we present an experimental study of the wrinkling threshold --- this is the experimentally observable quantity that we have been able to best understand. In our model problem, the thin sheet is loaded by gradually increasing the tension ratio $\gamma$ beyond one. To induce circumferential wrinkling in thin PS films, a water droplet can be placed onto the film, leading to wrinkle formation due to the capillary force exerted by the surface tension at the air–water–PS contact line \citep{huang2007capillary,vella2010capillary,schroll2013capillary}. Similarly, wrinkling can also be induced by poking a thin film or shell \citep{vella2011wrinkling,box2017indentation,vella2018regimes,box2019dynamics,wang2023effects}. However, in these loading scenarios, the stress state is fixed once a droplet is applied, making it difficult to continuously increase the stress exerted on the film. To overcome this limitation, we follow the experimental setup described in \citet{pineirua2013capillary,paulsen2017geometry} and use a controlled differential surface tension to impose $\gamma$. This is achieved by confining a sheet to a water--air interface and using a Langmuir trough (left panel of Figure \ref{fig:experiment}) to control the concentration of surfactant outside the film (and hence the effective tension $T_{\rm out}$). The interface within the holes remains pure throughout, and hence the tension $T_{\rm in}$ remains at the interfacial tension of water throughout.

Bearing in mind that our theoretical model considers a free-standing sheet, whereas the experiments are performed with the sheet floating on a water substrate, we discuss the applicability of the experimental setup to the physical quantities of interest. Specifically, we focus on the onset of wrinkling and the spatial extent of the wrinkled region, both of which are equilibrium characteristics determined by the quasi-static deformation of the sheet. The water substrate acts on the sheet only through hydrostatic pressure and it transmits no in-plane traction to the sheet. However, for ultra-thin sheets, the onset of wrinkling is governed solely by the in-plane stress field. Moreover, the spatial extent of the wrinkled region can be described by tension-field theory, whose leading-order prediction is independent of the substrate response. In view of these facts, the effects of both water viscosity and hydrostatic pressure are neglected in the present model.

We briefly introduce the materials and loading strategy used in our experiments. PS sheets (Young's modulus $3.4~\mathrm{GPa}$, Poisson's ratio $0.34$, and thickness $166~\mathrm{nm}$) are fabricated on glass substrates by spin coating the dissolved PS powder (Goodfellow) in toluene (anhydrous, 99.8\%; Sigma-Aldrich) and then cut into a circular shape with two circular holes, as shown in the right panel of Figure~\ref{fig:experiment}. To precisely control the geometric dimensions, we use a custom-designed template manufactured by laser cutting. Four cases are considered, corresponding to $\zeta = 0.5, 1, 1.5$, and $3$. All samples have the same overall diameter of $46~\mathrm{mm}$ and a hole diameter of $6~\mathrm{mm}$. The film is peeled off from the glass substrate using a deep water-filled beaker, after which it floats on the water surface. The floating film is then transferred into the Langmuir trough using a Petri dish. 

At the beginning of each experiment, the trough contains deionized water, so that the PS film is subjected to uniform tensions $T_\mathrm{in}\approx72~\mathrm{mN/m}$ and $T_\mathrm{out}\approx72~\mathrm{mN/m}$, corresponding to the surface tension of clean water; $\gamma=1$ initially. We then add the insoluble surfactant DPPC (dipalmitoylphosphatidylcholine) to the trough region outside the film by spreading from a chlorophorm solution. As a result, the outer surface tension $T_\mathrm{out}$ decreases, and its value is measured using a Wilhelmy plate. Upon moving inwards the barrier in the trough, the available trough area is reduced and the DPPC concentration increases, thereby decreasing the tension $T_\mathrm{out}$ applied at the outer boundary. Yet the surface tension $T_{\mathrm{in}}$ within the two holes remains unchanged during barrier compression because DPPC is insoluble in the water subphase. Consequently, as the barrier is moved toward the film, the tension ratio $\gamma$ increases and the film is loaded. Meanwhile, $\gamma$ can be precisely regulated. 

\begin{figure}[htbp]
    \centering
    \includegraphics[width=0.55\linewidth]{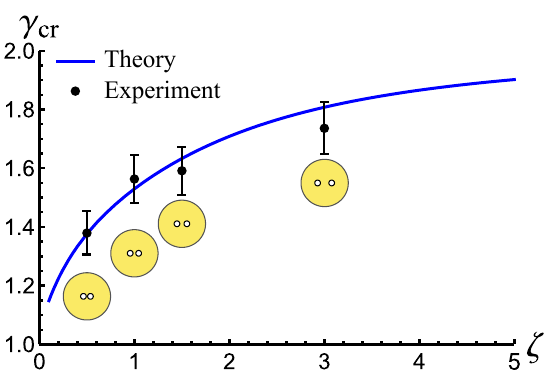}
    \caption{Comparison of the experimentally observed wrinkling threshold with the theoretical prediction. An error bar ($\pm5\%$) and the geometry of the associated sample are indicated.}
    \label{fig:comparison-threshold}
\end{figure}

We perform each experiment in the Langmuir trough and record the deformation process using a camera (Canon EOS 6D). The outer tension $T_\mathrm{out}$ is documented at a rate of one measurement per second whilst the barrier speed is $3\,\mathrm{cm/min}$. Guided by the near-threshold analysis in Section~\ref{near-threshold}, we  identify the onset of wrinkling around the hole and compare the experimental results with the theoretical prediction in Figure~\ref{fig:comparison-threshold}. The black dots correspond to the experimental data, with the sample geometry used in each experiment below the corresponding data point, while the solid curve represents the theoretical prediction based on the geometry with infinite size. Since tensional wrinkling is a supercritical bifurcation, it is not sensitive to imperfections. Additionally, the wrinkle amplitude grows gradually from zero. Taking this into account, we include a $\pm5\%$ error bar. The experimental results agree well with our theoretical predictions. We therefore infer that, during the deformation process, the effects of sheet thickness, liquid density, Young's modulus of the sheet, and gravity are negligible, as these factors do not enter the model for determining the threshold. These parameters, however, are known to affect the wrinkle number and amplitude \citep{coman2007wrinkling,davidovitch2011prototypical,pineirua2013capillary,schroll2013capillary}.

\begin{figure}[htbp]
    \centering
    \includegraphics[width=0.9\linewidth]{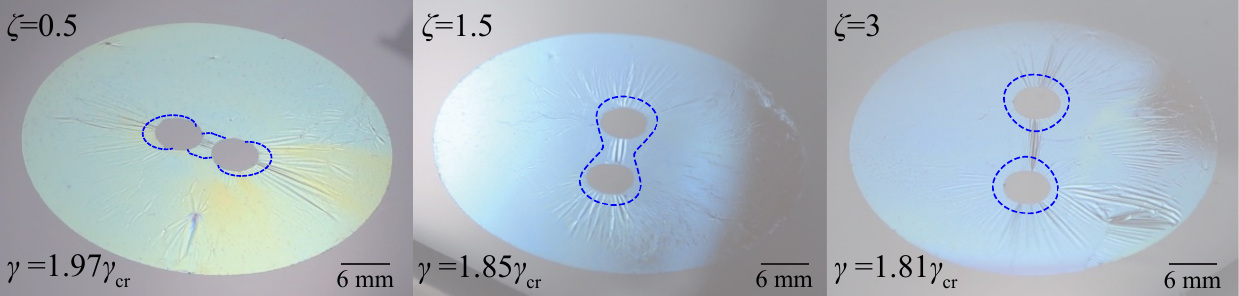}
    \caption{Snapshots of the wrinkled morphology as the tension ratio $\gamma$ exceeds the near-threshold regime. The critical tension ratio $\gamma_\mathrm{cr}$ takes the values $1.3765$, $1.6331$, and $1.8079$ in the left, middle, and right panels, respectively. For comparison, the boundary of the wrinkled region predicted by the near-threshold analysis is shown as a blue dashed line in each panel.}
    \label{fig:large-wrinkles}
\end{figure}

Finally, we present three snapshots of the wrinkled state in Figure~\ref{fig:large-wrinkles} for tension ratios beyond the near-threshold regime. The distance ratio and the corresponding tension ratio are given in each panel. In all cases, the tension ratio remains below $2\gamma_\mathrm{in}$, and wrinkles are observed to connect the two holes in all panels, even for $\zeta=3$. For comparison, the boundary of the wrinkled region predicted by the near-threshold analysis outlined in Section \ref{near-threshold} is also highlighted by a blue dashed line. It is observed that the actual extent of the wrinkled regions on the back sides is significantly larger than that predicted by the near-threshold theory. We mention that such an underestimation is also seen in the Lam\'{e} problem \citep{davidovitch2011prototypical}. Nevertheless, the near-threshold prediction captures qualitatively several  features of the wrinkle extent, including the overall shape of the wrinkled region and the direction along which the longest wrinkles develop. In particular, wrinkles are localized both in the central region between the two holes and in the back region. We note that, in both the Lam\'{e} problem and the annular geometry, wrinkles are uniformly distributed around the hole \citep{huang2007capillary,pineirua2013capillary,paulsen2017geometry,box2019dynamics}. In contrast, the presence of an additional hole breaks the radial symmetry, and the maximum wrinkle extent is observed along the direction connecting the centers of the two holes. This direction can therefore be identified as the preferred, or principal, direction of wrinkling. Furthermore, as discussed in \citet{davidovitch2011prototypical}, the wrinkle extent $W_{\rm FT}$ in the far-from-threshold regime of the Lam\'{e} problem reads $W_{\rm FT}=\gamma R_\mathrm{in}/2$. In the bipolar Lam\'{e} problem considered here, a rough estimate from Figure~\ref{fig:large-wrinkles} shows that the maximum $W_{\rm FT}$ is much larger than $\gamma R_\mathrm{in}/2$ ($\approx\gamma_\mathrm{cr}R_\mathrm{in}$ in all three cases), indicating that deformation or stress information can be transmitted over a longer distance and along a preferred direction.

\section{Conclusions}\label{conclusion}

We have studied the tension-induced wrinkling in thin elastic sheets containing two identical circular holes. We focused on thin sheets that have negligible bending stiffness and cannot support compressive stresses. An analytical solution to the pre-buckled state was obtained using bipolar coordinates. As a result, we derived the critical tension ratio $\gamma_\mathrm{cr}$ beyond which wrinkling is expected to take place. We further performed asymptotic analysis of the case in which the two holes are well-separated, obtaining excellent approximations for  $\gamma_\mathrm{cr}$. Near-threshold wrinkling was analyzed in detail to primarily unravel the basic feature of wrinkling initiation and evolution. We find that wrinkles first occur at the  area between the two holes and then gradually expand to the back sides. Further, we derived a perturbative solution for the far-from-threshold scenario and obtained qualitative information concerning the wrinkle distribution.

To test our predictions, we carried out experimental investigations using a Langmuir trough to generate wrinkling in a circular PS sheet that contains two circular holes. The  predicted critical tension ratio $\gamma_\mathrm{cr}$ was  validated.  Experimental snapshots of the wrinkle morphology in the regime where the tension ratio $\gamma$ exceeds the critical threshold are presented, providing novel insight into the wrinkle distribution. In comparison with the far-from-threshold results for the Lam\'{e} problem, we find that a much larger wrinkle extent is achieved experimentally. Specifically, the breaking of radial symmetry induces a preferred (or principal) direction along which the wrinkle extent attains its maximum reach. We expect that the results presented in this paper may also shed new light on wrinkling phenomena in thin films induced by multiple surface bubbles \citep{oratis2020new}. We also note that the wrinkles shown in Figure~\ref{fig:large-wrinkles} with $\gamma \sim 2\gamma_\mathrm{in}$ extend close to the outer circular edge of the sheet. When these wrinkles reach the edge, we would expect that other deformation modes become available to the sheet, including folding \citep{paulsen2017geometry}.

Tension-induced wrinkling in  elastic films is not only of fundamental interest  in thin-film mechanics but also has  significant potential for evaluating forces generated by cell traction \citep{cerda2003geometry,style2014traction,li2022wrinkle}. Additionally, growing evidence show that mechanical cues can be sensed by cells to regulate their migration, differentiation, and collective cell behaviors \citep{engler2006matrix,ladoux2017mechanobiology,vining2017mechanical,petridou2017multiscale,murthy2017piezos,van2018mechanoreciprocity}. In this context, the Lam\'{e} problem provides an instructive analogy for understanding the situation in which a single cell contracts on a compliant substrate. A remarkable result is that the stress distribution can be regulated by wrinkling, such that the maximum stress decays inversely with distance rather than with its square, thereby enabling force transmission over longer ranges \citep{davidovitch2011prototypical}. It is well known that cells can respond to long-range signals generated by other cells \citep{alisafaei2021long}. For the case of two cells, our model offers preliminary insights on the critical tension needed to generate wrinkles and their distributions. 

More broadly, this study contributes to a growing perspective in solid mechanics that views wrinkling not merely as a failure mode or imperfection, but as a functional response that allows thin structures to adapt to complex loading and environmental conditions. Whether in engineered membranes, soft matter systems, or biological tissues, tension-induced wrinkling emerges as a powerful mechanism for stress redistribution and long-range mechanical communication.

Finally, we emphasize that the primary objective of this work is to develop a fundamental understanding of how the presence of an additional hole affects both the onset of wrinkling and the spatial extent of the wrinkled region. Other important factors, such as finite bending stiffness, unequal hole sizes, and non-circular hole geometries, may significantly influence both the onset and the distribution of wrinkles. For example, when the bending stiffness cannot be neglected, the simple stress criterion adopted in the present work is no longer sufficient, and a rigorous stability analysis is required to
determine the onset of wrinkling. In such cases, finite element simulations or other numerical approaches may provide effective alternatives. Moreover, the post-buckling characteristics, including the wrinkle wavelength and amplitude, are beyond the scope of the present study. These quantities are closely related to both the bending stiffness of the sheet and the out-of-plane normal stress \citep{davidovitch2011prototypical,pineirua2013capillary,Ardaseva2026}. More comprehensive theoretical and numerical models incorporating these effects will be developed in future work.



\section*{Acknowledgment}
 This work was supported by a grant from the National Natural Science Foundation of China (Project No. 12372072). Y.L.~and A.G.~acknowledge the UKRI Horizon Europe Guarantee MSCA (Marie Sk\l odowska-Curie Actions) Postdoctoral Fellowship (EPSRC Grant No. EP/Y030559/1). S.R.~acknowledges the Wolfson Visiting Fellowship, which provided support during her sabbatical visit to the University of Cambridge. We thank Dr. Aleksandra Arda\v{s}eva for helpful discussions at the early stages of this project, and Dr.~Ben Fudge at University of Oxford for assisting with the laser cutting. Y.L.~also thanks Dr.~Ming Dai at Nanjing University of Aeronautics and Astronautics for insightful discussions of stress divergence when two holes are extremely close.
 

\appendix
\section{Standard results for a bipolar coordinate system}\label{appendix:bipolar}

In this Appendix, we briefly summarize some standard results expressed in bipolar coordinates.

\subsection{Coordinate system}

Let $\{\e_\tau,\e_\xi\}$ be the bipolar orthonormal basis  while $\{\e_1,\e_2\}$ is  the Cartesian one, then we have
\begin{equation}
\e_\tau=\frac{1-\cos\xi\cosh\tau}{\cosh\tau-\cos\xi}\e_1-\frac{\sin\xi\sinh\tau}{\cosh\tau-\cos\xi}\e_2,\quad \e_\xi=-\frac{\sin\xi\sinh\tau}{\cosh\tau-\cos\xi}\e_1+\frac{\cos\xi\cosh\tau-1}{\cosh\tau-\cos\xi}\e_2,\label{eq:base}
\end{equation}
and, conversely,
\begin{equation}
\e_1=\frac{1-\cos\xi\cosh\tau}{\cosh\tau-\cos\xi}\e_\tau+\frac{\sin\xi\sinh\tau}{\cosh\tau-\cos\xi}\e_\xi,\quad \e_2=-\frac{\sin\xi\sinh\tau}{\cosh\tau-\cos\xi}\e_\tau+\frac{\cos\xi\cosh\tau-1}{\cosh\tau-\cos\xi}\e_\xi.
\end{equation}
From \eqref{eq:base}, we obtain
\begin{equation}
\frac{\partial\e_\tau}{\partial\tau}=\frac{h\sin\xi}{a}\e_\xi,\quad\frac{\partial\e_\tau}{\partial\xi}=-\frac{h\sinh\tau}{a}\e_\xi,\quad\frac{\partial\e_\xi}{\partial\tau}=-\frac{h\sin\xi}{a}\e_\tau,\quad\frac{\partial\e_\xi}{\partial\xi}=\frac{h\sinh\tau}{a}\e_\tau.
\label{Dbase}
\end{equation}

\subsection{Stress tensor and equations of equilibrium}
In the plane, we can write the Cauchy stress tensor with respect to the bipolar basis as
\begin{equation}
\bm\sigma=\sigma_{\tau\tau}\e_\tau\otimes\e_\tau+\sigma_{\tau\xi}\e_\tau\otimes\e_\xi+\sigma_{\tau\xi}\e_\xi\otimes\e_\tau+\sigma_{\xi\xi}\e_\xi\otimes\e_\xi.
\label{eq:cauchy-stress}
\end{equation}
Using  the definition of divergence operator in terms of a general orthonormal coordinate system \citep[see p.~268 of][for example]{goriely17} and the scale factor \eqref{eq:scalor}, we arrive at 
\begin{equation}
\begin{aligned}
&\frac{1}{h}\frac{\partial\sigma_{\tau\tau}}{\partial\tau}+\frac{1}{h}\frac{\partial\sigma_{\tau\xi}}{\partial\xi}+\frac{\sinh\tau}{a}\left(\sigma_{\xi\xi}-\sigma_{\tau\tau}\right)-\frac{2\sin\xi}{a}\sigma_{\tau\xi}=0,\\
&\frac{1}{h}\frac{\partial\sigma_{\xi\xi}}{\partial\xi}+\frac{1}{h}\frac{\partial\sigma_{\tau\xi}}{\partial\tau}+\frac{\sin\xi}{a}\left(\sigma_{\tau\tau}-\sigma_{\xi\xi}\right)-\frac{2\sinh\tau}{a}\sigma_{\tau\xi}=0.
\end{aligned}\label{eq:equilibrium}
\end{equation}

The strain components $\varepsilon_{ij}$ in bipolar coordinates are:
\begin{equation}
    \begin{aligned}
        &\varepsilon_{\tau\tau}=\frac{1}{h}\frac{\partial u}{\partial \tau}-\frac{v\sin\xi}{a},\quad  \varepsilon_{\xi\xi}=\frac{1}{h}\frac{\partial v}{\partial \xi}-\frac{u\sinh\tau}{a},\\&
         \varepsilon_{\tau\xi}=\frac{1}{2}\left(\frac{1}{h}\frac{\partial v}{\partial \tau}+\frac{1}{h}\frac{\partial u}{\partial \xi}+\frac{u\sin\xi}{a}+\frac{v\sinh\tau}{a}\right),
    \end{aligned}
\end{equation}
where $u(\tau,\xi)$ and $v(\tau,\xi)$ are displacement components in $\xi$- and $\tau$-directions, respectively.

\subsection{Airy stress function}
In the bipolar geometry used here the stress components are given in terms of $\psi$ by \citep{jeffery1921ix,ling1948stresses,wu1970general,lucht2015bipolar}
\begin{align}
  \sigma_{\tau\tau}&=\frac{1}{a}\left((\cosh\tau-\cos\xi)\frac{\partial^2}{\partial\xi^2}-\sinh\tau\frac{\partial}{\partial\tau}-\sin\xi\frac{\partial}{\partial\xi}+\cosh\tau\right)\frac{\psi}{h},\label{eq:sigmatautau}\\  \sigma_{\xi\xi}&=\frac{1}{a}\left((\cosh\tau-\cos\xi)\frac{\partial^2}{\partial\tau^2}-\sinh\tau\frac{\partial}{\partial\tau}-\sin\xi\frac{\partial}{\partial\xi}+\cos\xi\right)\frac{\psi}{h},\label{eq:sigmaxixi}\\  \sigma_{\tau\xi}&=-\frac{1}{a}\left(\cosh\tau-\cos\xi\right)\frac{\partial^2}{\partial\tau\partial\xi}\frac{\psi}{h}.\label{eq:sigmatauxi}
\end{align}

\section{Derivation of the approximate stresses in the far-distance limit of two holes}\label{appendixa}

In this Appendix, we derive the asymptotic expressions for the stress components in the limit $\zeta \to \infty$. Referring to Figure~\ref{fig:bipolar}, we identify
\begin{equation}
    \quad\xi=\theta_1-\theta_2,\quad d_1^2=r_1^2+r_2^2-2r_1r_2\cos\xi,\quad 
\end{equation}
where the second relation follows directly from the law of cosines. Moreover, from $\eqref{eq:radius}_1$ and \eqref{eq:tauin}, we obtain
\begin{equation}
    a=R_\mathrm{in}\sinh\tau_\mathrm{in}=\frac{\sqrt{\zeta(\zeta+4)}}{2}R_\mathrm{in}.
    \label{eq:a-zeta}
\end{equation}
To proceed further, we introduce the following scaling for large $\zeta$:
\begin{equation}
    r_1\sim \frac{a}{\sinh\tau},
    \label{eq:r1}
\end{equation}
where $\tau$ denotes the bipolar coordinate of the material point shown in Figure~\ref{fig:bipolar}.

It is expected that as $\zeta\rightarrow\infty$, $\sigma_{\tau\tau}$ will reduce to the $\sigma_{rr}$ (the poloar coordinates are used for the classic Lam\'{e} problem). Furthermore, we make use of the identities 
\begin{equation}
  \begin{aligned}
       &r_2^2=r_1^2+d_1^2+2r_1d_1\cos\theta_1,\quad \cos\xi=\frac{r_1+d_1\cos\theta_1}{r_2},\\&
       \sin\xi=\frac{d_1\sin\theta_1}{r_2},\quad \cos2\xi=\frac{r_1^2+2r_1d_1\cos\theta_1+d_1^2\cos2\theta_1}{r_2^2},
  \end{aligned}
\end{equation}
and then take the series expansions of \eqref{eq:sigmatautau1}-\eqref{eq:sigmatauxi1} to obtain
\begin{equation}
    \begin{aligned}
        &\sigma_{\tau\tau}=T_\mathrm{out}+(T_\mathrm{in}-T_\mathrm{out})\frac{R_\mathrm{in}^2}{r_1^2}+\frac{2(T_\mathrm{in}-T_\mathrm{out})\cos\theta_1}{\zeta}\frac{R_\mathrm{in}^3}{r_1^3}+O\left(\zeta^{-2}\right),\\
        &\sigma_{\tau\xi}=2(T_\mathrm{in}-T_\mathrm{out})\frac{R_\mathrm{in}\left(r_1^2-R_\mathrm{in}^2\right)}{r_1^3\zeta}\sin\theta_1+O\left(\zeta^{-2}\right),\\
        &\sigma_{\xi\xi}=T_\mathrm{out}-(T_\mathrm{in}-T_\mathrm{out})\frac{R_\mathrm{in}^2}{r_1^2}-\frac{2(T_\mathrm{in}-T_\mathrm{out})\cos\theta_1}{\zeta}\frac{R_\mathrm{in}^3}{r_1^3}+O\left(\zeta^{-2}\right).
    \end{aligned}\label{eq:stress-far}
\end{equation}
It can be seen that, as $\zeta\rightarrow \infty$, stresses in \eqref{eq:stress-far} fully recover the classical Lam\'{e} solution \citep{timoshenko,coman2007wrinkling,davidovitch2011prototypical}. We point out that the terms of $O\left(\zeta^{-2}\right)$ are lengthy so we omit them here.

According to \eqref{eq:stress-far} with terms of $O\left(\zeta^{-2}\right)$ being involved, we can also determine the threshold $\gamma_\mathrm{cr}$. To this end, we write down the coordinate $r_1$ at the inner boundary, and it is actually related to $R_\mathrm{in}$ by
\begin{equation}
    r_1^2+a^2\left(\coth\tau_\mathrm{in}-1\right)^2-2r_1a\left(\coth\tau_\mathrm{in}-1\right)\cos\theta_1=R_\mathrm{in}^2.
\end{equation}
Combining the fact that $R_\mathrm{in}=a/\sinh\tau_\mathrm{in}$ and $\tau_\mathrm{in}=\operatorname{arccosh(\zeta/2+1)}$, we denote the coordinate at the inner boundary in terms of $\theta_1$ and $\zeta$. Finally, from the equation $\sigma_2=0$ we derive 
\begin{equation}
    \gamma=2-\frac{4\cos2\theta_1}{\zeta^2}+\frac{2\left(2\cos\theta_1+8\cos2\theta_1+2\cos3\theta_1\right)}{\zeta^3}+O\left(\zeta^{-4}\right),
\end{equation}
which attains a minimum at $\theta_1=\pi$ with the associated critical tension ratio given by
\begin{equation}
    \gamma_\mathrm{cr}=2-\frac{4}{\zeta^2}+\frac{8}{\zeta^3}+O\left(\zeta^{-3}\right),
\end{equation}
which is consistent with \eqref{eq:asy-gamma-four} up to $O\left(\zeta^{-3}\right)$. 
This validates the consistency of our asymptotic analysis.

\bibliographystyle{elsarticle-harv}
\bibliography{references}

\end{document}